\documentclass[pra,onecolumn,floatfix,a4paper,superscriptaddress]{revtex4}
\usepackage{bm,color,graphicx,amsmath,txfonts}
\usepackage{here}
\usepackage{array}
\usepackage{graphicx}
\usepackage[colorlinks, citecolor=blue,linkcolor=blue]{hyperref}
\usepackage{braket}
\usepackage{dsfont}
\usepackage[left=	2cm,top=2cm,right=1.2cm,bottom=2cm]{geometry}
\usepackage{tikz}
\everymath{\displaystyle}

\begin{document}
\makeatletter
\renewcommand{\@biblabel}[1]{%
	\makebox[2.3em][l]{\fontsize{10}{13}\selectfont[#1]}}
\makeatother

\title{Multiparameter quantum estimation and Stirling Engine Performance in a Gravitational Cat State System}

\author{Omar Bachain}
\address{LPHE-Modeling and Simulation, Faculty of Sciences, Mohammed V University in Rabat, Rabat, Morocco}

\author{Mohamed \surname{Amazioug} }
\email{m.amazioug@uiz.ac.ma}
\address{LPTHE-Department of Physics, Faculty of Sciences, Ibnou Zohr University, Agadir 80000, Morocco}

\author{Rachid Ahl Laamara}
\address{LPHE-Modeling and Simulation, Faculty of Sciences, Mohammed V University in Rabat, Rabat, Morocco}
\address{Centre of Physics and Mathematics, CPM, Faculty of Sciences, Mohammed V University in Rabat, Rabat, Morocco}

\date{\today}
\begin{abstract}

	We investigate the multiparameter quantum estimation and quantum thermodynamics properties of a gravitational cat state (gravcat) system composed of two interacting massive particles confined in double-well potentials. The system is described by an effective Hamiltonian involving the energy splitting parameter $\omega$ and the gravitational coupling strength $\gamma$, while the interaction with a thermal environment is modeled through a Gibbs thermal state. Within the framework of quantum parameter estimation theory, we employ the quantum Fisher information matrix (QFIM) to analyze the precision limits for estimating the three fundamental parameters of the model, namely the gravitational coupling $\gamma$, the energy splitting $\omega$, and the temperature $T$. Utilizing the symmetric logarithmic derivative (SLD) formalism within the QFIM framework, we derive the analytical expressions of the estimation bounds and evaluate the corresponding minimal variances associated with the quantum Cram\'er-Rao bound. Both simultaneous and individual estimation strategies are investigated, and their performances are compared in different parameter regimes. Our results reveal the existence of optimal estimation regions where the precision is significantly enhanced and show that the relative efficiency of the estimation schemes strongly depends on the interaction strength, the energy gap, and the thermal environment. In addition, the thermodynamic behavior of the system is analyzed within the framework of a quantum Stirling cycle. The internal energy, entropy, heat exchanges, and work production are examined, allowing us to evaluate the efficiency of the gravcat-based quantum heat engine. The obtained results highlight the interplay between quantum metrology and quantum thermodynamics.
\end{abstract}
\maketitle

\section{Introduction}

Quantum metrology has become one of the most important research
directions in quantum information science due to its ability to enhance
the precision of parameter estimation beyond classical limits \cite{pirandola2018advances,Giovannetti2011,mukhopadhyay2025current}. The
central idea of quantum metrology is to exploit quantum resources such
as entanglement, coherence, and quantum correlations in order to
improve measurement sensitivity. In this context, the concept of Fisher
information plays a fundamental role since it quantifies the amount of
information that a quantum state carries about an unknown parameter.
Originally introduced in classical statistics, Fisher information is
directly related to the Cramér–Rao inequality \cite{Safranek2018,Mondal2025,Genoni2013,Asjad2023}, which establishes a
lower bound on the variance of any unbiased estimator
\cite{Fisher1925,Cramer1946,Rao1945}. The extension of this concept to
the quantum regime has led to the development of quantum Fisher
information (QFI), which determines the ultimate precision limits
allowed by quantum mechanics in parameter estimation problems
\cite{Helstrom1976,Holevo1982}.

A fundamental result in quantum estimation theory is the quantum
Cramér–Rao bound (QCRB) \cite{Cavazzoni2025,He2025,Sidhu2020,Demkowicz2020}, which states that the variance of any unbiased
estimator is bounded by the inverse of the quantum Fisher information.
Consequently, maximizing the QFI is equivalent to minimizing the
estimation uncertainty. This concept has found numerous applications in
quantum technologies, including quantum sensing \cite{Montenegro2025}, quantum imaging,
atomic interferometry, and gravitational wave detection
\cite{Braunstein1994,Giovannetti2011,DemkowiczDobrzanski2015}. In recent
years, increasing attention has been devoted to multiparameter quantum
estimation \cite{Paris2009,Gessner2023,Gebhart2024,AbdRabbou2022}, where several parameters characterizing a quantum system
are estimated simultaneously \cite{Bachain2026}. In such situations, the relevant quantity
is the quantum Fisher information matrix (QFIM), whose inverse
determines the lower bound of the covariance matrix associated with the
estimated parameters \cite{Liu2019,Ragy2016}. However, the simultaneous
estimation of multiple parameters introduces additional challenges
related to the possible incompatibility between optimal measurements
for different parameters \cite{Alanazi2025}. For this reason, the comparison between
simultaneous and individual estimation strategies has become an
important problem in quantum metrology
\cite{Alipour2014}.

Parallel to these developments, recent advances in quantum physics have
stimulated significant interest in the study of macroscopic quantum
superpositions and their possible implications for fundamental
physics. In particular, the investigation of massive quantum systems
prepared in spatial superposition states has opened new perspectives
for testing the interplay between quantum mechanics and gravity. One
of the most intriguing proposals in this direction is the concept of
gravitational cat states (gravcats), which correspond to massive
particles placed in spatial superpositions that generate distinct
gravitational fields \cite{Bose2017,Marletto2017}. These proposals have
triggered extensive theoretical and experimental efforts aimed at
probing the quantum nature of gravity through the observation of
gravity-induced entanglement between massive systems. Such systems can
be effectively modeled as interacting two-level systems whose dynamics
are governed by an effective Hamiltonian that incorporates both
gravitational interactions and internal energy splitting.

In addition to their fundamental interest, gravcat systems provide a
promising platform for exploring the interplay between quantum
information theory, quantum metrology, and quantum thermodynamics \cite{Partovi1989,Gemmer2009,Utreras2015,Abah2018}.
Quantum thermodynamics has emerged as an active research field devoted
to understanding how thermodynamic concepts such as work \cite{Amazioug2024}, heat, and
entropy behave at the quantum scale. In particular, quantum heat
engines based on microscopic systems have attracted considerable
attention because they provide a natural framework for studying energy
conversion processes governed by quantum laws \cite{Su2016,Carnot1872,Kosloff2014,Feldmann2004,Rezek2006}. Among the different
models of quantum heat engines, the Stirling cycle has been widely
investigated due to its conceptual simplicity and its relevance for
finite-time thermodynamics \cite{Bachain2025}. The study of such engines allows one to
analyze how quantum effects influence thermodynamic quantities such as
efficiency, work production, and heat exchang \cite{Berry2006,Abliz2006,Li2008,Souza2008,Ahadpour2021,Zhang2007,Park2020,Li2008thermal,Xi2017}.

Motivated by these developments, the present work investigates the
quantum metrological and thermodynamic properties of a gravitational
cat state system composed of two interacting massive particles
confined in double-well potentials \cite{Rojas2023,Lobo2025}. The system is described by an
effective Hamiltonian involving the gravitational coupling strength
$\gamma$ and the energy splitting parameter $\omega$, while the
interaction with a thermal environment is taken into account through
a Gibbs thermal state characterized by the temperature $T$. Using the
quantum Fisher information matrix formalism, we analyze the ultimate
precision limits for estimating the three fundamental parameters of
the model, namely the gravitational interaction strength $\gamma$, the
energy splitting $\omega$, and the temperature $T$. Both simultaneous
and individual estimation strategies are examined in order to identify
the optimal parameter regimes that minimize the estimation uncertainty.

Furthermore, we investigate the thermodynamic behavior of the system
within the framework of a quantum Stirling cycle. By evaluating
thermodynamic quantities such as the von Neumann entropy, internal
energy, heat exchange, and work production, we analyze the performance
of the gravcat system as a quantum heat engine. The obtained results
provide new insights into the interplay between quantum metrology and
quantum thermodynamics and highlight the role of optimal parameter
estimation in improving the thermodynamic performance of quantum
devices.

The remainder of this paper is organized as follows. In Sec.~\ref{sec2}, we
introduce the physical model and derive the thermal density matrix of
the system. In Sec.~\ref{sec3}, we present the formalism of the quantum Fisher
information matrix used for multiparameter estimation. In Sec.~\ref{sec4}, we
analyze the estimation precision for different parameter pairs using
both simultaneous and individual strategies. In Sec.~\ref{sec5}, we investigate
the thermodynamic properties of the system within a quantum Stirling
cycle. In Sec.~\ref{sec6}, we discuss the feasibility and experimental
prospects of the proposed model. Finally, the main conclusions of this
work are summarized in Sec.~\ref{sec7}.

\section{Hamiltonian and thermal interaction
}\label{sec2}
The notion of gravitational cat states (gravcats) arises from the interplay between quantum superposition and gravitational interaction. In this framework, a quantum state initially localized in space can be coherently delocalized over a characteristic length scale $L$. Such a configuration relies on the possibility that a massive quantum system may exist in a spatial superposition of different gravitational potentials. This idea has been discussed in several theoretical works addressing macroscopic quantum coherence in gravitational contexts \cite{Penrose1996}. Moreover, it is closely related to recent studies investigating gravity-induced entanglement and nonlocal correlations in optomechanical systems and matter-wave interferometry \cite{Bose2017,Marletto2017}. At sufficiently small scales, a particle with mass $m$ may therefore occupy a coherent superposition of two spatially separated positions, in analogy with the Schrödinger cat state but with explicit gravitational implications.
The physical system considered here consists of two massive particles confined in independent one-dimensional double-well potentials. Each particle can be effectively described as a two-level system (qubit), where the two minima of the potential correspond to localized states $|+\rangle$ and $|-\rangle$. These states satisfy
\begin{equation}
	\hat{x}|\pm\rangle = \pm \frac{L}{2} |\pm\rangle ,\label{cond}
\end{equation}
where $L$ denotes the separation between the two potential minima. The corresponding Hamiltonian of a single particle reads
\begin{equation}
	\hat{H}=\frac{\hat{p}^{2}}{2m}+U(\hat{x}),
\end{equation}
with $\hat{p}$ the momentum operator and $U(\hat{x})$ the double-well potential. The localized states can be expressed in terms of the energy eigenstates as
\begin{equation}
	|\pm\rangle = \frac{1}{\sqrt{2}}\left(|0\rangle \pm |1\rangle\right),
\end{equation}
where $|0\rangle$ and $|1\rangle$ denote the ground and first excited states of the particle, respectively, and $\omega$ represents the corresponding energy splitting.
The validity of the condition (\ref{cond}) requires that decoherence effects remain sufficiently weak to preserve spatial coherence. In particular, the coherence time of the system should exceed the characteristic time scales associated with gravitational interaction and quantum tunneling. Experimental investigations have indicated that under cryogenic temperatures and ultra-high vacuum conditions, environmental dephasing can be significantly suppressed, thereby maintaining the coherence of the superposition state \cite{RomeroIsart2011,Hornberger2012}. Furthermore, theoretical studies suggest that gravitationally induced decoherence becomes negligible for masses below a certain threshold, which supports the applicability of this model to mesoscopic quantum systems \cite{Diosi1984}.
In the non-relativistic regime, the effective Hamiltonian describing the gravitational interaction between two gravcats can be written as \cite{Anastopoulos2020}
\begin{equation}
\mathcal{H}=\frac{\omega}{2}\left(\sigma_z\otimes I + I\otimes\sigma_z\right)-\gamma(\sigma_x\otimes\sigma_x),
	\label{Hgravcat}
\end{equation}
where $I$ denotes the identity operator and $\sigma_x$ and $\sigma_z$ are the Pauli matrices acting on the internal states of each gravcat. The parameter $\gamma$ quantifies the strength of the gravitational coupling and is defined as
\begin{equation}
	\gamma=\frac{Gm^2}{2}\left(\frac{1}{d}-\frac{1}{d^\prime}\right),
\end{equation}
with $G$ the gravitational constant and $d$ and $d^\prime$ representing the relative separations between the two particles. These distances correspond to specific configurations where the gravitational forces acting on each gravcat are minimized.
In the computational basis $\{|00\rangle,|01\rangle,|10\rangle,|11\rangle\}$, the Hamiltonian in Eq.~(\ref{Hgravcat}) can be represented by the matrix
\begin{equation}
	\mathcal{H}=
	\begin{pmatrix}
		\omega & 0 & 0 & -\gamma \\
		0 & 0 & -\gamma & 0 \\
		0 & -\gamma & 0 & 0 \\
		-\gamma & 0 & 0 & -\omega
	\end{pmatrix}.\label{Hamil}
\end{equation}
Diagonalizing this Hamiltonian leads to the eigenvalues
\begin{equation}
	E_{1,2} = \mp \gamma,
	\qquad
	E_{3,4} = \mp \sqrt{\omega^2 + \gamma^2},\label{vap}
\end{equation}
with the corresponding eigenstates
\begin{align}
	|\psi_1\rangle &= \frac{1}{\sqrt{2}}(|01\rangle + |10\rangle), \\
	|\psi_2\rangle &= \frac{1}{\sqrt{2}}(|01\rangle - |10\rangle), \\
	|\psi_3\rangle &= \cos(\nu^+) |00\rangle + \sin(\nu^+) |11\rangle, \\
	|\psi_4\rangle &= \cos(\nu^-) |00\rangle + \sin(\nu^-) |11\rangle,
\end{align}
where
\begin{equation}
	\nu^\pm = \arctan\left(
	\frac{\gamma}{\omega \pm \sqrt{\omega^2+\gamma^2}}
	\right).
\end{equation}
\begin{figure}[H]
	\centering
	\includegraphics[width=0.4\linewidth,
		trim=2cm 10cm 2cm 4cm,
	clip]{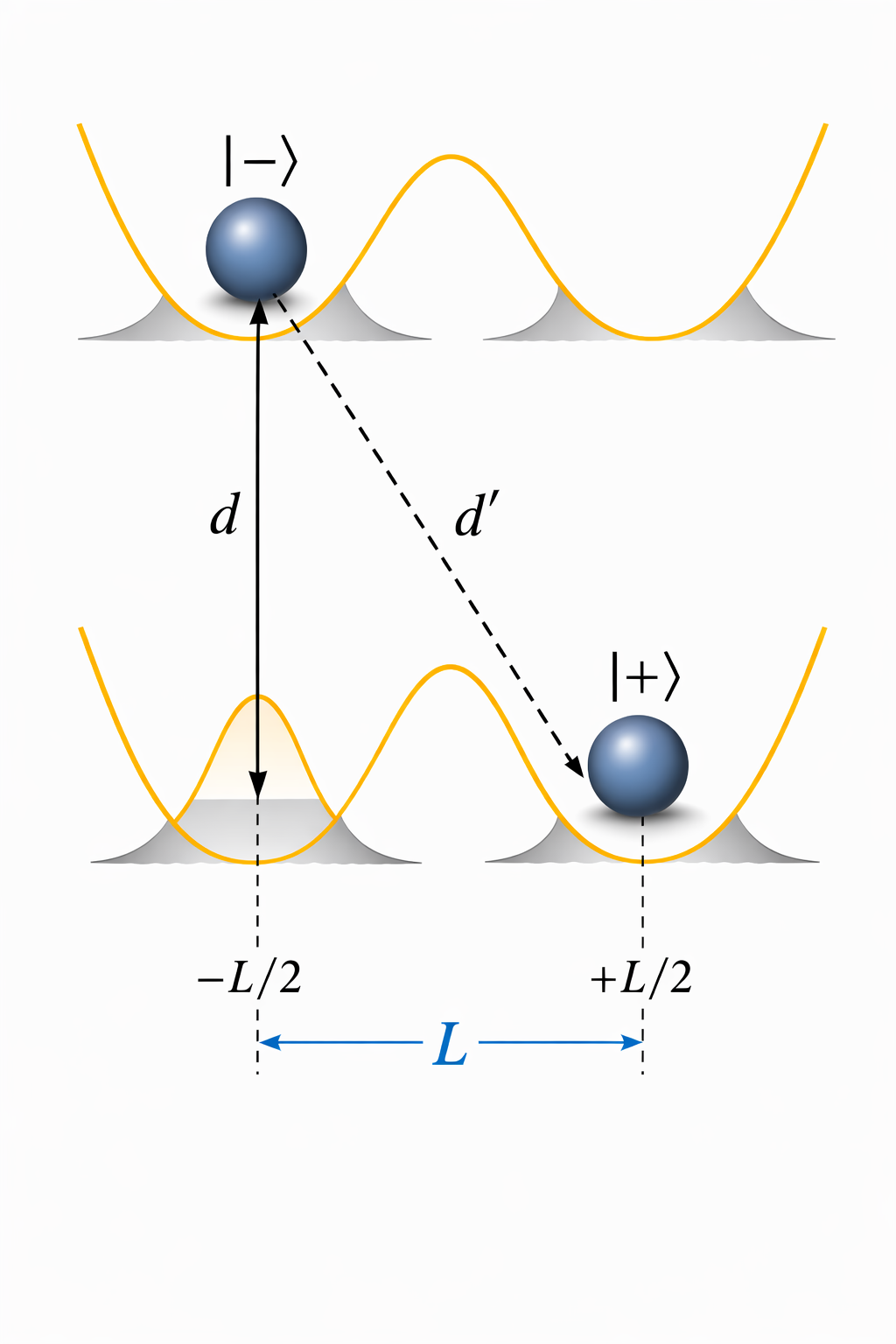}
	\caption{
		Schematic diagram of the gravcats model. Each symmetric double-well potential is arranged 
		along a distinct axis, while two of these axes are parallel and separated by a distance 
		$d=\sqrt{(d^{'})^2-L^2}$.
	}
\end{figure}
In realistic physical situations, the quantum system is typically in contact with an external environment. 
Consequently, the equilibrium state of the system at finite temperature is described by the Gibbs density operator. 
Assuming that the system reaches thermal equilibrium with a heat bath at temperature $T$, the density matrix is given by
\begin{equation}
	\varrho(T)=\frac{e^{-\beta \mathcal{H}}}{Z},
\end{equation}
where $\mathcal{H}$ is the Hamiltonian defined in Eq.~(\ref{Hamil}), 
$\beta=1/(k_B T)$ denotes the inverse temperature, $k_B$ is the Boltzmann constant (here we set $k_B=1$ for simplicity), 
and $Z$ represents the partition function defined as
\begin{equation}
	Z=\text{Tr}\left[e^{-\beta \mathcal{H}}\right].
\end{equation}

Using the eigenvalues of the Hamiltonian, the partition function can be written as

\begin{equation}
	Z=2\cosh(\beta\gamma)+2\cosh\left(\beta\sqrt{\omega^2+\gamma^2}\right).
\end{equation}

In the computational basis $\{|00\rangle,|01\rangle,|10\rangle,|11\rangle\}$, 
the thermal density matrix takes the so-called X-state form

\begin{equation}
	\varrho^T=
	\begin{pmatrix}
		x & 0 & 0 & \eta \\
		0 & z & \delta & 0 \\
		0 & \delta & z & 0 \\
		\eta & 0 & 0 & y
	\end{pmatrix},\label{rho}
\end{equation}

where the non-zero elements are expressed as

\begin{align}\nonumber
	x &= \frac{\cos^2(\nu^+)e^{-\beta\,E_3}+\cos^2(\nu^-)e^{-\beta\,E_4}}{Z}, \\\nonumber
	z &= \frac{e^{-\beta\,E_1}+e^{-\beta\,E_2}}{2Z}, \\
	\delta &=\frac{e^{-\beta\,E_1}-e^{-\beta\,E_2}}{2Z}, \\
	\eta &= \frac{\sin(2\nu^+)e^{-\beta\,E_3}+\sin(2\nu^-)e^{-\beta\,E_4}}{Z},\nonumber\\
	y &= \frac{\sin^2(\nu^+)e^{-\beta\,E_3}+\sin^2(\nu^-)e^{-\beta\,E_4}}{Z},\nonumber
\end{align}

where $\epsilon_i$ are the eigenvalues of the Hamiltonian and $\nu^{\pm}$ are defined by

\begin{equation}
	\nu^\pm=\arctan\left(\frac{\gamma}{\omega\pm\sqrt{\omega^2+\gamma^2}}\right).
\end{equation}

It is worth noting that the X-structure of the density matrix originates from the symmetry of the Hamiltonian. 
The thermal density operator $\varrho^T$ satisfies the standard conditions of a density matrix, namely 
$\text{Tr}\left( \varrho^T\right) =1$ and $\text{Tr}\left[ \left( \varrho^T\right) ^2\right] \leq1$. 

\section{Quantum Fisher information matrix (QFIM)}\label{sec3}

The Quantum Fisher Information Matrix (QFIM) plays a fundamental role in
quantum parameter estimation as it determines the ultimate precision
limits allowed by quantum mechanics. In general, calculating the QFIM
requires the diagonalization of the density matrix. However, an
alternative and more convenient formulation exists for finite
dimensional systems that avoids the explicit diagonalization procedure.
Let $B \in M_{n\times n}$ be an $n \times n$ matrix. The vectorization
operator $\mathrm{vec}(\cdot)$ is defined as \cite{gilchrist2009vectorization}

\begin{equation}
	\mathrm{vec}[B] =
	(b_{11},\ldots,b_{n1},b_{12},b_{22},\ldots,b_{1n},\ldots,b_{nn})^{T}.\label{vec}
\end{equation}

For the estimation of a single parameter $\xi$, the Quantum Fisher
Information (QFI) is given by 

\begin{equation}
	\mathcal{F}(\varrho_\xi)=\mathrm{Tr}\{\varrho_\xi \mathcal{L}_\xi^{2}\},
\end{equation}

where $\mathcal{L}_\xi$ denotes the symmetric logarithmic derivative (SLD)
operator.
When several parameters $\{\xi_i\}$ are simultaneously estimated,
the relevant quantity becomes the Quantum Fisher Information Matrix
whose elements are defined as \cite{paris2009quantum}

\begin{equation}
	\mathcal{F}_{ij}=\frac{1}{2}\mathrm{Tr}\left[
	(\mathcal{L}_{\xi_i}\mathcal{L}_{\xi_j}+L_{\xi_j}\mathcal{L}_{\xi_i})\varrho
	\right].
\end{equation}

The SLD operators satisfy the relation

\begin{equation}
	2\,\partial_{\xi_i}\varrho = \mathcal{L}_{\xi_i}\varrho + \varrho \mathcal{L}_{\xi_i}.
\end{equation}

If the density matrix admits the spectral decomposition
$\varrho=\sum_m p_m |m\rangle\langle m|$, the QFIM elements can be written
as \cite{banchi2014quantum,sommers2003bures}

\begin{equation}
	\mathcal{F}_{ij}=2\sum_{p_m+p_n>0}
	\frac{\langle m|\partial_{\xi_i}\varrho|n\rangle
		\langle n|\partial_{\xi_j}\varrho|m\rangle}
	{p_m+p_n}.
\end{equation}

Correspondingly, the SLD operator takes the form

\begin{equation}
	\mathcal{L}_{\xi_i}=2\sum_{p_m+p_n>0}
	\frac{\langle m|\partial_{\xi_i}\varrho|n\rangle}
	{p_m+p_n}|m\rangle\langle n|.
\end{equation}

Another representation of the QFIM can be expressed through the
integral form \cite{paris2009quantum}

\begin{equation}
	\mathcal{F}_{ij}=2\int_{0}^{\infty}
	\mathrm{Tr}\left[
	e^{-\varrho t}(\partial_{\xi_i}\varrho)
	e^{-\varrho t}(\partial_{\xi_j}\varrho)
	\right]dt.
\end{equation}

A more practical explicit expression was introduced using the
vectorization technique of the density matrix. In this approach the matrix \cite{vsafranek2018simple}

\begin{equation}
	\mathcal{R} = (\varrho^{T}\otimes I + I \otimes \varrho),
\end{equation}

is defined, allowing the QFIM to be written as

\begin{equation}
\mathcal{F}_{ij}=
	\partial_{\xi_i}\mathrm{vec}[\varrho]^{\dagger}
	\mathcal{R}^{-1}
	\partial_{\xi_j}\mathrm{vec}[\varrho].
\end{equation}

Similarly, the vectorized form of the SLD operator becomes

\begin{equation}
	\mathrm{vec}[L_{\xi_i}]
	=2\mathcal{R}^{-1}\mathrm{vec}[\partial_{\xi_i}\varrho].\label{SLD}
\end{equation}

The fundamental relationship between the QFIM and the precision limit
in quantum estimation theory is given by the Quantum Cram\'er-Rao Bound (QCRB) 

\begin{equation}
	\mathrm{Cov}(\hat{\xi}) \geq \mathcal{F}^{-1},
\end{equation}

where $\mathrm{Cov}(\hat{\xi})$ denotes the covariance matrix of the
estimated parameters.
For the single–parameter case, the QCRB reduces to

\begin{equation}
	\mathrm{Var}(\hat{\xi}_i) \geq (\mathcal{F}^{-1})_{ii}.
\end{equation}

This bound can be saturated when the optimal measurements are
constructed from the eigenvectors of the SLD operator. In the
multiparameter scenario, however, achieving the bound is generally more
challenging. Although the commutation condition

\begin{equation}
	[\mathcal{L}_{\xi_i},\mathcal{L}_{\xi_j}]=0
\end{equation}

is sufficient to guarantee the simultaneous attainability of the QCRB,
it is not strictly necessary. A weaker condition ensuring the
possibility of saturating the bound is \cite{ragy2016compatibility,napoli2019towards,carollo2018uhlmann,adani2024critical}

\begin{equation}
	\mathrm{Tr}\big(\varrho [\mathcal{L}_{\xi_i},\mathcal{L}_{\xi_j}]\big)=0 .
\end{equation}
\section{Results and Discussion}\label{sec4}
In this section, we derive the explicit form of the Quantum Fisher Information Matrix (QFIM), which plays a central role in quantum parameter estimation theory. This quantity provides the ultimate bound on the precision with which unknown parameters characterizing a quantum system can be estimated. In particular, we employ the QFIM framework to analyze the estimation of the gravitational interaction strength $\gamma$ between the two gravcats, the energy difference $\omega$ between the relevant quantum states, and the temperature $T$ of the considered thermal environment. By evaluating the corresponding elements of the QFIM, we investigate how these physical parameters influence the estimation precision and determine the optimal conditions for achieving high-precision measurements.
\subsection{Estimation of parameters $\gamma$ and $T$}

In this section, we investigate the feasibility of estimating the two
parameters $\gamma$ and $\omega$ using the Quantum Fisher Information
Matrix (QFIM). The estimation procedure relies on the vectorization
method applied to the derivatives of the density matrix with respect
to the considered parameters.
Using the definition of the vec-operator given in Eq.~(\ref{vec}), the
vectorized derivatives of the density matrix with respect to $\gamma$
and $\omega$ can be written as

\begin{equation}
	\mathrm{vec}[\partial_\gamma \varrho]=
	[\partial_\gamma x,0,0,\partial_\gamma \eta,0,\partial_\gamma z,\partial_\gamma \delta,0,0,\partial_\gamma \delta,\partial_\gamma z,0,\partial_\gamma \eta,0,0,\partial_\gamma y]^T ,\label{vecg}
\end{equation}

\begin{equation}
	\mathrm{vec}[\partial_T \varrho]=
[\partial_T x,0,0,\partial_T \eta,0,\partial_T z,\partial_T \delta,0,0,\partial_T \delta,\partial_T z,0,\partial_T \eta,0,0,\partial_T y]^T , .
\end{equation}

Accordingly, the quantum Fisher information matrix can be expressed as

\begin{equation}
	\mathcal{F}=
	\begin{pmatrix}
		\mathcal{F}_{\gamma\gamma} & \mathcal{F}_{\gamma T} \\
		\mathcal{F}_{T\gamma} & \mathcal{F}_{T T}
	\end{pmatrix}
	=
	\begin{pmatrix}
		2\,\mathrm{vec}[\partial_\gamma\varrho]^\dagger
		\mathcal{R}^{-1}\mathrm{vec}[\partial_\gamma\varrho] &
		2\,\mathrm{vec}[\partial_\gamma\varrho]^\dagger
		\mathcal{R}^{-1}\mathrm{vec}[\partial_T\varrho] \\
		2\,\mathrm{vec}[\partial_T\varrho]^\dagger
		\mathcal{R}^{-1}\mathrm{vec}[\partial_\gamma\varrho] &
		2\,\mathrm{vec}[\partial_T\varrho]^\dagger
		\mathcal{R}^{-1}\mathrm{vec}[\partial_T\varrho]
	\end{pmatrix},
\end{equation}
The analytical expressions of the elements of the Quantum Fisher
Information Matrix for the estimation of the parameters $(\gamma,T)$
are rather lengthy and are therefore reported in Appendix \ref{Appendix A}.
The ultimate precision achievable in quantum parameter estimation is
bounded by the Quantum Cramér–Rao bound (QCRB). For the parameters
$\hat{\xi}\equiv(\gamma,T)$, the covariance matrix satisfies

\begin{equation}
	\mathrm{Cov}(\hat{\xi}) \ge \mathcal{F}^{-1}.
\end{equation}

The inverse of the QFIM is given by

\begin{equation}
	\mathcal{F}^{-1}=
	\frac{1}{\det(\mathcal{F})}
	\begin{pmatrix}
		\mathcal{F}_{TT} & -\mathcal{F}_{ T \gamma} \\
		-\mathcal{F}_{\gamma T} & \mathcal{F}_{\gamma\gamma}
	\end{pmatrix}.
\end{equation}

From the inequality above, one obtains the following bounds for the
variances of the estimators

\begin{equation}
	\mathrm{Var}(\gamma)\ge
	\frac{\mathcal{F}_{TT}}{\det(\mathcal{F})},\qquad
	\mathrm{Var}(T)\ge
	\frac{\mathcal{F}_{\gamma\gamma}}{\det(\mathcal{F})}.
\end{equation}

Furthermore, the covariance relation between the two parameters reads

\begin{equation}
	\left(\mathrm{Var}(\gamma)-\frac{\mathcal{F}_{TT}}{\det(F)}\right)
	\left(\mathrm{Var}(T)-\frac{\mathcal{F}_{\gamma\gamma}}{\det(F)}\right)
	\ge
	\left(\mathrm{Cov}(\gamma,T)+
	\frac{\mathcal{F}_{\gamma T}}{\det(\mathcal{F})}\right)^2 .
\end{equation}

Using Eq.~(\ref{SLD}), the symmetric logarithmic derivative (SLD) operators
associated with the parameters $\gamma$ and $\omega$ can be written as

\begin{equation}
	\mathcal{L}_\gamma=
	\begin{pmatrix}
			\mathcal{L}_{11}^\gamma & 0 & 0 & 	\mathcal{L}_{14}^\gamma \\
		0 & 	\mathcal{L}_{22}^\gamma & \mathcal{L}_{21}^\gamma & 0 \\
		0 & \mathcal{L}_{32}^\gamma & 	\mathcal{L}_{33}^\gamma & 0 \\
			\mathcal{L}_{41}^\gamma & 0 & 0 & 	\mathcal{L}_{44}^\gamma
	\end{pmatrix},
	\qquad
	L_T=
	\begin{pmatrix}
	\mathcal{L}_{11}^T & 0 & 0 & 	\mathcal{L}_{14}^T \\
	0 & 	\mathcal{L}_{22}^T & \mathcal{L}_{21}^T & 0 \\
	0 & \mathcal{L}_{32}^T & 	\mathcal{L}_{33}^T & 0 \\
	\mathcal{L}_{41}^T & 0 & 0 & 	\mathcal{L}_{44}^T
	\end{pmatrix}.\label{Lg}
\end{equation}

The elements of these operators are expressed as

\begin{align}\nonumber
	\mathcal{L}_{11}^\gamma &=\frac{\eta ^2 (\partial_\gamma y-\partial_\gamma x)+x(x+y)\,\partial_\gamma x  -2\eta  y\, \partial_\gamma \eta }{(x+y) \left(x y-\eta ^2\right)} , \\\nonumber
		\mathcal{L}_{22}^\gamma &=	\mathcal{L}_{3,3}^\gamma
	=\frac{ z\,\partial_\gamma z-\delta \,\partial_\gamma \delta}{z^2-\delta ^2}, \\
		\mathcal{L}_{14}^\gamma &=	\mathcal{L}_{41}^\gamma=\frac{ \eta  y\,\partial_\gamma x+ \eta  x \,\partial_\gamma y-2  x y\,\partial_\gamma \eta}{(x+y) \left(\eta ^2-x y\right)} ,\\\nonumber
	\mathcal{L}_{23}^\gamma &=	\mathcal{L}_{32}^\gamma= 	\frac{ z\,\partial_\gamma \delta-\delta \, \partial_\gamma z}{z^2-\delta ^2}\\\nonumber
	\mathcal{L}_{44}^\gamma &=\frac{\eta ^2
	(\partial_\gamma x-\partial_\gamma y)+ x (x+y)\,\partial_\gamma y-2  \eta  x\,\partial_\gamma \eta}{(x+y) \left(x y-\eta ^2\right)}\nonumber
\end{align}

and similarly for the parameter $\omega$

\begin{align}\nonumber
	\mathcal{L}_{11}^T &=\frac{\eta ^2 (\partial_T y-\partial_T x)+x(x+y)\,\partial_T x  -2\eta  y\, \partial_T \eta }{(x+y) \left(x y-\eta ^2\right)} , \\\nonumber
\mathcal{L}_{22}^T &=	\mathcal{L}_{3,3}^T
=\frac{ z\,\partial_T z-\delta \,\partial_T \delta}{z^2-\delta ^2}, \\
\mathcal{L}_{14}^T &=	\mathcal{L}_{41}^T=\frac{ \eta  y\,\partial_T x+ \eta  x \,\partial_T y-2  x y\,\partial_T \eta}{(x+y) \left(\eta ^2-x y\right)} ,\\\nonumber
\mathcal{L}_{23}^T &=	\mathcal{L}_{32}^T= 	\frac{ z\,\partial_T \delta-\delta \, \partial_T z}{z^2-\delta ^2}\\\nonumber
\mathcal{L}_{44}^T &=\frac{\eta ^2
	(\partial_T x-\partial_T y)+ x (x+y)\,\partial_Ty-2  \eta  x\,\partial_T \eta}{(x+y) \left(x y-\eta ^2\right)}\nonumber
\end{align}

Finally, by saturating the QCRB inequalities, the minimum achievable
variances for the estimation of the parameters $\gamma$ and $T$
are obtained as

\begin{equation}
	\mathrm{Var}(\gamma)_{\min}=
	\frac{\mathcal{F}_{TT}}{\det(\mathcal{F})},\qquad
	\mathrm{Var}(\omega)_{\min}=
	\frac{\mathcal{F}_{\gamma\gamma}}{\det(\mathcal{F})} .
\end{equation}

\begin{figure}[H]
	\centering
	\includegraphics[scale=0.8]{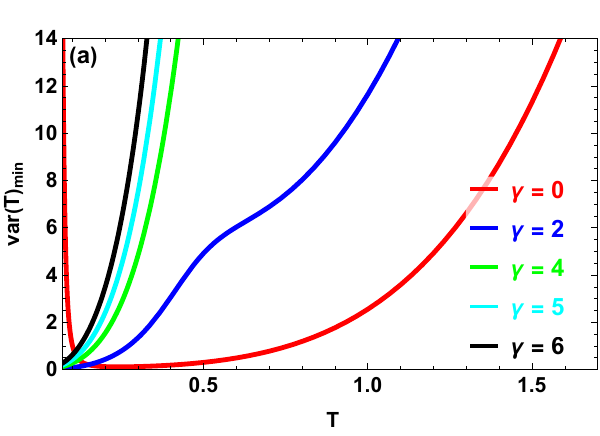}
	\hspace*{0.3cm}
	\includegraphics[scale=0.8]{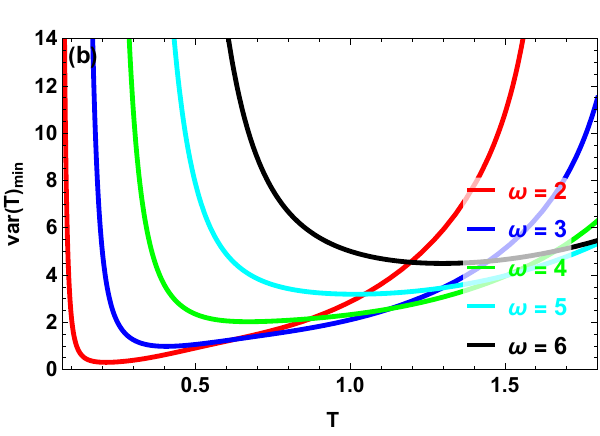}
	\caption{Minimal variance for the simultaneous estimation of the temperature $T$ as a function of $T$: 
		(a) for different values of $\gamma$ with $\omega = 1$; 
		(b) for different values of $\omega$ with $\gamma = 1.5$.}
	\label{fig:1}
\end{figure}
Figure \ref{fig:1}(a) shows the variation of the minimal variance $\mathrm{Var}(T)_{\min}$ as a function of the temperature $T$ for different values of the gravitational coupling parameter $\gamma$. It can be observed that the estimation precision strongly depends on both the temperature and the interaction strength. At low temperatures, the minimal variance remains very small, indicating a high precision in temperature estimation. As the temperature increases, $\mathrm{Var}(T)_{\min}$ gradually grows, which reflects a decrease in the estimation accuracy due to the increased thermal population of the excited energy levels. Moreover, increasing the coupling parameter $\gamma$ significantly modifies the behavior of the curves and shifts the region of optimal precision toward lower temperatures. This behavior originates from the modification of the energy spectrum induced by the gravitational interaction.

Figure \ref{fig:1}(b) presents the minimal variance $\mathrm{Var}(T)_{\min}$ as a function of $T$ for different values of the energy splitting parameter $\omega$. In this case, a clear minimum appears in each curve, indicating the existence of an optimal temperature at which the estimation precision is maximized. Furthermore, increasing $\omega$ shifts the optimal estimation region toward higher temperatures. This behavior can be attributed to the enlargement of the energy gap between the system levels, which requires higher thermal energy to significantly populate the excited states. These results demonstrate that the estimation precision of the temperature can be effectively controlled by tuning the physical parameters $\gamma$ and $\omega$ of the gravcats system.
\begin{figure}[H]
	\centering
	\includegraphics[scale=0.8]{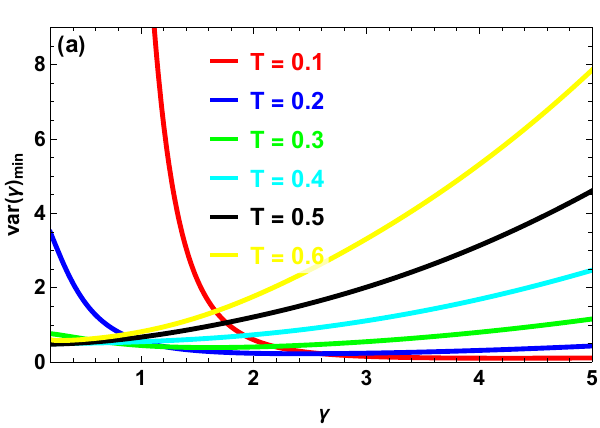}
	\hspace*{0.3cm}
	\includegraphics[scale=0.8]{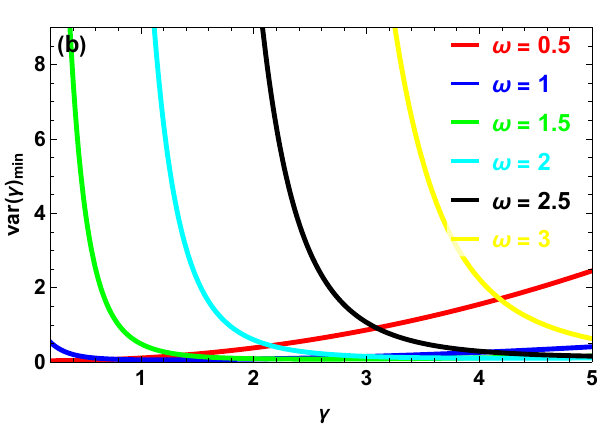}
	\caption{Minimal variance for the simultaneous estimation of $\gamma$ as a function of $\gamma$: 
		(a) for different values of $T$ with $\omega = 2$; 
		(b) for different values of $\omega$ with $T = 0.1$.}
	\label{fig:2}
\end{figure}
Figure \ref{fig:2}(a) shows the variation of the minimal variance $\mathrm{Var}(\gamma)_{\min}$ as a function of the gravitational coupling parameter $\gamma$ for different values of the temperature $T$. It can be observed that the estimation precision strongly depends on both $\gamma$ and the thermal environment. For very small values of $\gamma$, the variance becomes relatively large for some temperatures, indicating a lower estimation precision. As $\gamma$ increases, the minimal variance rapidly decreases and reaches very small values, showing that the estimation of the gravitational coupling becomes more accurate in the strong interaction regime. Moreover, increasing the temperature generally leads to larger variances, which reflects the detrimental effect of thermal fluctuations on the estimation precision.

Figure \ref{fig:2}(b) illustrates $\mathrm{Var}(\gamma)_{\min}$ as a function of $\gamma$ for different values of the energy splitting parameter $\omega$. The results indicate that the estimation precision is highly sensitive to the energy gap of the system. For small values of $\omega$, the minimal variance remains relatively low over a broad range of $\gamma$, implying a better estimation performance. In contrast, increasing $\omega$ significantly modifies the behavior of the curves and shifts the optimal estimation region toward larger values of $\gamma$. This behavior originates from the modification of the system energy spectrum, which directly affects the quantum Fisher information and therefore the achievable precision in the estimation of the gravitational coupling parameter.

The Cramér--Rao inequality for the individual estimation of the parameters can be written as
\begin{equation}
	\mathrm{Var}(\gamma)^{\mathrm{Ind}} \geq \mathcal{F}^{-1}_{\gamma\gamma}, \qquad
	\mathrm{Var}(T)^{\mathrm{Ind}} \geq \mathcal{F}^{-1}_{TT},
	\label{eq:CR}
\end{equation}
where $\mathcal{F}^{-1}$ denotes the inverse of the Fisher information matrix. 
The corresponding minimal variances, which define the ultimate precision bounds, are therefore given by
\begin{equation}
	\mathrm{Var}(\gamma)^{\mathrm{Ind}}_{\min} = \mathcal{F}^{-1}_{\gamma\gamma}, \qquad
	\mathrm{Var}(T)^{\mathrm{Ind}}_{\min} = \mathcal{F}^{-1}_{TT}.
\end{equation}
\begin{figure}[H]
	\centering
	\includegraphics[scale=0.8]{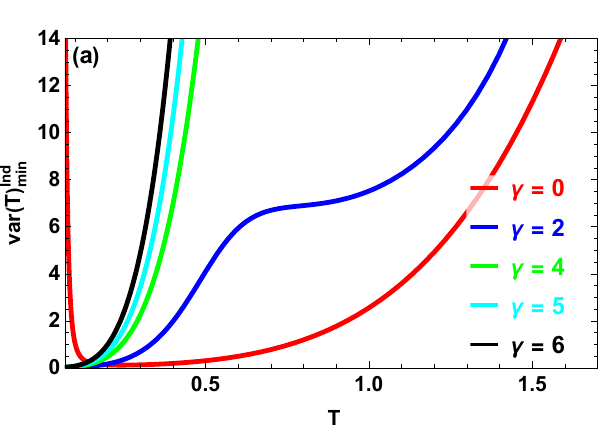}
	\hspace*{0.3cm}
	\includegraphics[scale=0.8]{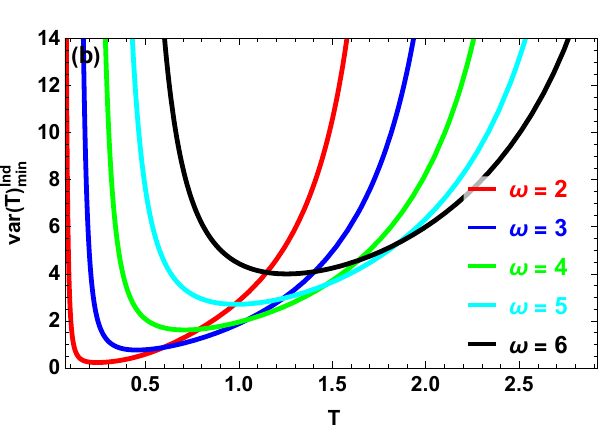}
	\caption{Minimal variance for the  individual estimation of $T$ as a function of $T$: 
		(a) for different values of $\gamma$ with $\omega = 1$; 
		(b) for different values of $\omega$ with $\gamma = 1.5$.}
	\label{fig:3}
\end{figure}
Figures \ref{fig:3}(a) and \ref{fig:3}(b) show the minimal variance associated with the individual estimation of the temperature $T$ as a function of $T$. In Fig. \ref{fig:3}(a), the results are plotted for different values of the gravitational coupling parameter $\gamma$ with a fixed energy splitting $\omega=1$. As shown in the figure, the minimal variance remains very small at low temperatures, indicating a high precision in the estimation of $T$. However, as the temperature increases, the variance gradually grows, which reflects the reduction of the estimation accuracy due to the increasing thermal population of the excited states. Furthermore, increasing the coupling strength $\gamma$ significantly modifies the behavior of the curves and shifts the region of optimal precision toward lower temperatures.
Figure \ref{fig:3}(b) presents the same quantity for different values of the energy splitting parameter $\omega$ with a fixed coupling $\gamma=1.5$. The curves exhibit a clear minimum, indicating the existence of an optimal temperature at which the estimation precision becomes maximal. Moreover, increasing $\omega$ shifts this optimal region toward higher temperatures. This behavior can be attributed to the enlargement of the energy gap of the system, which requires higher thermal energy to significantly populate the excited levels and thus affects the sensitivity of the system to temperature variations.
\begin{figure}[H]
	\centering
	\includegraphics[scale=0.8]{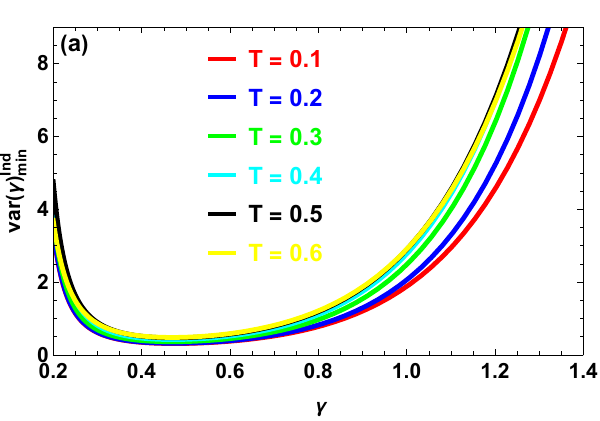}
	\hspace*{0.3cm}
	\includegraphics[scale=0.8]{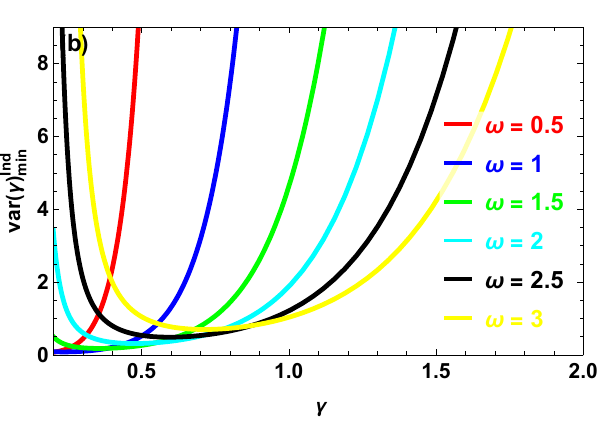}
	\caption{Minimal variance for the individual estimation of $\gamma$ as a function of $\gamma$: 
		(a) for different values of $T$ with $\omega = 2$; 
		(b) for different values of $\omega$ with $T = 0.1$.}
	\label{fig:4}
\end{figure}
The minimal variance associated with the individual estimation of the gravitational coupling parameter $\gamma$ is plotted as a function of $\gamma$ for several values of the temperature $T$ with $\omega = 2$, as illustrated in Fig.~\ref{fig:4}(a). The curves exhibit a pronounced minimum, indicating the presence of an optimal interaction strength where the estimation precision reaches its highest level. For very small values of $\gamma$, the variance remains relatively large, while it decreases as $\gamma$ approaches the optimal region. Beyond this point, the variance increases again due to the reduced sensitivity of the quantum state to further variations of the coupling parameter. Moreover, increasing the temperature slightly enlarges the minimal variance, revealing the detrimental effect of thermal fluctuations on the estimation accuracy.

The dependence of $\mathrm{Var}(\gamma_{\mathrm{Ind}})_{\min}$ on $\gamma$ for different values of the energy splitting parameter $\omega$ at $T=0.1$ is presented in Fig.~\ref{fig:4}(b). A clear shift in the optimal estimation region is observed when $\omega$ varies. In particular, increasing the energy splitting modifies the curvature of the variance profiles and changes the value of $\gamma$ that minimizes the variance. This behavior can be attributed to the modification of the system energy spectrum, which directly affects the sensitivity of the quantum Fisher information to variations of the gravitational coupling parameter.

The ratio between the total variances of the two estimation schemes is defined as
\begin{equation}
	\Gamma(\gamma,T)=\frac{\Delta(\gamma,T)_{\mathrm{Sim}}}{\Delta(\gamma,T)_{\mathrm{Ind}}}
	=\frac
	{\frac{1}{2}\left(\mathrm{Var}(\gamma)_{\min}+\mathrm{Var}(T)_{\min}\right)}{\left(\mathrm{Var}(\gamma)_{\min}^{\mathrm{Ind}}+\mathrm{Var}(T)_{\min}^{\mathrm{Ind}}\right)} .
\end{equation}

When $\Gamma(\gamma,T)< 1$, i.e., $\Delta_{\mathrm{Sim}}<\Delta_{\mathrm{Ind}}$, the simultaneous estimation strategy provides a higher precision than the individual estimation scheme. Conversely, when $\Gamma(\gamma,T)>1$, meaning $\Delta_{\mathrm{Sim}}>\Delta_{\mathrm{Ind}}$, the individual estimation becomes more efficient in improving the precision.
\begin{figure}[H]
	\centering
	\includegraphics[scale=0.8]{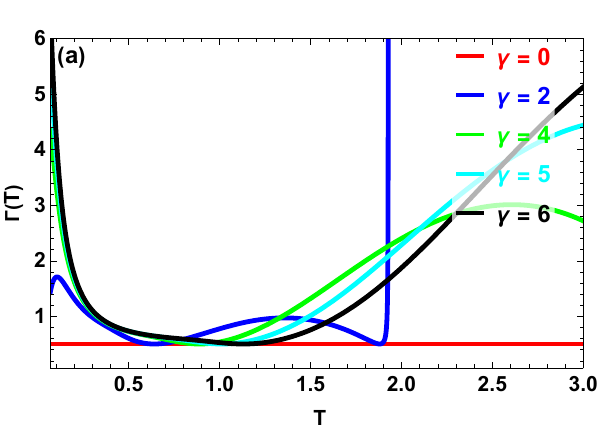}
	\hspace*{0.3cm}
	\includegraphics[scale=0.8]{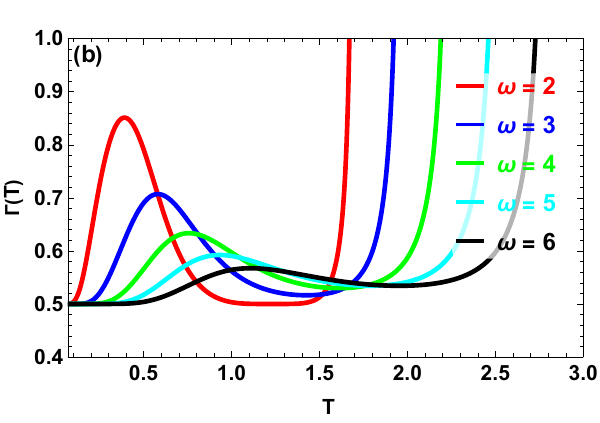}
	\caption{The ratio between the minimal total variances in the estimation of the parameters $T$ and $\gamma$ as a function of $T$: 
		(a) for different values of $\gamma$ with $\omega = 1$; 
		(b) for different values of $\omega$ with $\gamma = 1.5$.}
	\label{fig:5}
\end{figure}
The temperature dependence of $\Gamma(T)$ for different values of the coupling parameter $\gamma$ with $\omega=1$ is displayed in Fig.~\ref{fig:5}(a). In the low–temperature regime, the ratio remains below unity for most values of $\gamma$, indicating that the simultaneous estimation scheme provides a higher precision than the individual estimation strategy. As the temperature increases, $\Gamma(T)$ can exceed one for certain interaction strengths, revealing a crossover region where the individual estimation becomes more advantageous. This behavior highlights the important role played by thermal effects in determining the relative performance of the two estimation schemes. 

The influence of the energy splitting parameter $\omega$ on $\Gamma(T)$ at fixed $\gamma=1.5$ is illustrated in Fig.~\ref{fig:5}(b). Over a wide range of temperatures, the ratio remains smaller than unity, showing that the simultaneous estimation protocol generally outperforms the individual estimation. Increasing $\omega$ slightly modifies the temperature interval where the advantage of the simultaneous strategy is most pronounced, which can be attributed to the modification of the system energy spectrum and the corresponding redistribution of thermal populations.
\begin{figure}[H]
	\centering
	\includegraphics[scale=0.8]{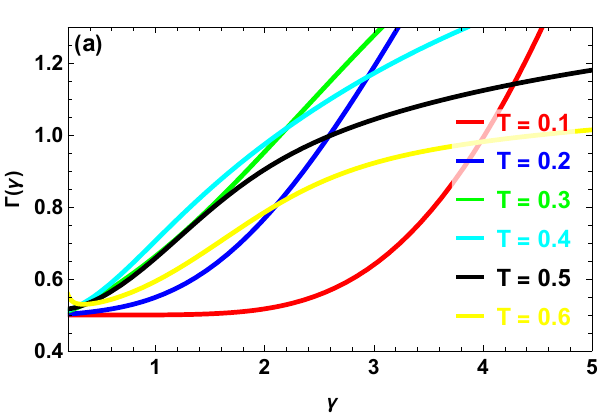}
	\hspace*{0.3cm}
	\includegraphics[scale=0.8]{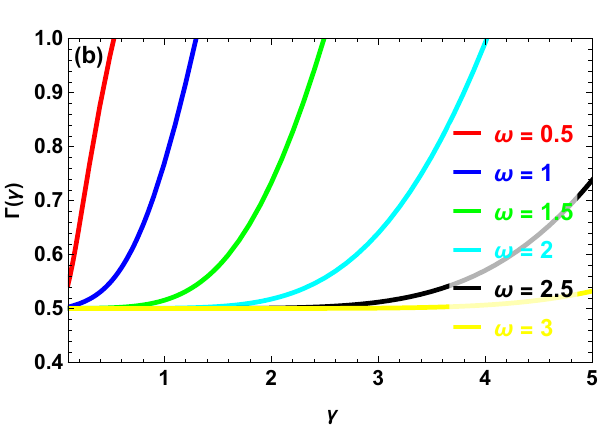}
	\caption{The ratio between the minimal total variances in the estimation of the parameters $T$ and $\gamma$ as a function of $\gamma$: 
		(a) for different values of $T$ with $\omega = 2$; 
		(b) for different values of $\omega$ with $T = 0.1$.}
	\label{fig:6}
\end{figure}
The dependence of $\Gamma(\gamma)$ on the coupling parameter $\gamma$ for different values of the temperature $T$ with $\omega=2$ is illustrated in Fig.~\ref{fig:6}(a). For small interaction strengths, the ratio remains below unity for most temperatures, indicating that the simultaneous estimation strategy provides a better precision than the individual estimation scheme. As $\gamma$ increases, $\Gamma(\gamma)$ progressively grows and may exceed one for certain temperatures, revealing a transition region where the individual estimation becomes comparatively more efficient. This behavior highlights the significant role played by both the interaction strength and the thermal fluctuations in determining the relative performance of the two estimation protocols. 

The influence of the energy splitting parameter $\omega$ on $\Gamma(\gamma)$ at fixed $T=0.1$ is presented in Fig.~\ref{fig:6}(b). Over a broad range of the coupling parameter $\gamma$, the ratio generally remains below unity, which indicates that the simultaneous estimation scheme maintains an advantage over the individual strategy. Increasing $\omega$ slightly modifies the dependence of $\Gamma(\gamma)$ on $\gamma$, reflecting the impact of the energy spectrum structure on the attainable precision in the multiparameter estimation process. 

\subsection{Estimation of parameters $\omega$ and $T$}

In this subsection, we investigate the joint estimation of the 
parameter $\omega$ and the temperature $T$. 
Following the same procedure adopted in the previous subsection,
the derivatives of the density matrix with respect to these parameters
are first evaluated and expressed using the vectorization technique.
According to the definition of the $\mathrm{vec}$ operator, the
corresponding vectors can be written as

\begin{equation}
	\mathrm{vec}[\partial_\omega \varrho]=
	[\partial_\omega x,0,0,\partial_\omega \eta,0,\partial_\omega z,\partial_\omega \delta,0,0,\partial_\omega \delta,\partial_\omega z,0,\partial_\omega \eta,0,0,\partial_\omega y]^T ,\label{veco}
\end{equation}

\begin{equation}
	\mathrm{vec}[\partial_T \varrho]=
	[\partial_T x,0,0,\partial_T \eta,0,\partial_T z,\partial_T \delta,0,0,\partial_T \delta,\partial_T z,0,\partial_T \eta,0,0,\partial_T y]^T.
\end{equation}

Using these expressions together with the vectorization formalism of
the density matrix, the quantum Fisher information matrix (QFIM)
associated with the parameters $(\omega,T)$ can be obtained as

\begin{equation}
	\mathcal{F}=
	\begin{pmatrix}
		\mathcal{F}_{\omega\omega} & \mathcal{F}_{\omega T} \\
		\mathcal{F}_{T\omega} & \mathcal{F}_{T T}
	\end{pmatrix}
	=
	\begin{pmatrix}
		2\,\mathrm{vec}[\partial_\omega\varrho]^\dagger
		\mathcal{R}^{-1}\mathrm{vec}[\partial_\omega\varrho] &
		2\,\mathrm{vec}[\partial_\omega\varrho]^\dagger
		\mathcal{R}^{-1}\mathrm{vec}[\partial_T\varrho] \\
		2\,\mathrm{vec}[\partial_T\varrho]^\dagger
		\mathcal{R}^{-1}\mathrm{vec}[\partial_\omega\varrho] &
		2\,\mathrm{vec}[\partial_T\varrho]^\dagger
		\mathcal{R}^{-1}\mathrm{vec}[\partial_T\varrho]
	\end{pmatrix}.
\end{equation}
Due to their lengthy form, the analytical expressions of the Quantum Fisher Information Matrix elements corresponding to the estimation of $(\omega,T)$ are reported in Appendix \ref{Appendix A}.
The ultimate precision achievable in estimating the parameters
$\omega$ and $T$ is constrained by the quantum
Cramér--Rao bound (QCRB). Accordingly, the variances of the estimators
satisfy the following inequalities

\begin{equation}
	\mathrm{Var}(\omega) \geq \frac{\mathcal{F}_{TT}}{\det(\mathcal{F})},
	\qquad
	\mathrm{Var}(T) \geq \frac{\mathcal{F}_{\omega\omega}}{\det(\mathcal{F})},
\end{equation}

\begin{equation}
	\left(
	\mathrm{Var}(\omega)-\frac{\mathcal{F}_{TT}}{\det(\mathcal{F})}
	\right)
	\left(
	\mathrm{Var}(T)-\frac{\mathcal{F}_{\omega\omega}}{\det(\mathcal{F})}
	\right)
	\geq
	\left(
	\mathrm{Cov}(\omega,T)+\frac{\mathcal{F}_{\omega T}}{\det(\mathcal{F})}
	\right)^2 .
\end{equation}

The symmetric logarithmic derivative (SLD) operators associated with
$\omega$ and $T$ can then be derived from the vectorized form,
leading to the operators $\mathcal{L}_{\omega}$ and $\mathcal{L}_{T}$, whose matrix
elements depend on the components of $\mathfrak{R}^{-1}$ and the
derivatives of the density matrix elements, which can be written in the matrix form as
\begin{equation}
	\mathcal{L}_\omega=
	\begin{pmatrix}
		\mathcal{L}_{11}^\omega & 0 & 0 & 	\mathcal{L}_{14}^\omega \\
		0 & 	\mathcal{L}_{22}^\omega & \mathcal{L}_{21}^\omega & 0 \\
		0 & \mathcal{L}_{32}^\omega & 	\mathcal{L}_{33}^\omega & 0 \\
		\mathcal{L}_{41}^\omega & 0 & 0 & 	\mathcal{L}_{44}^\omega
	\end{pmatrix},
	\qquad
	L_T=
	\begin{pmatrix}
		\mathcal{L}_{11}^T & 0 & 0 & 	\mathcal{L}_{14}^T \\
		0 & 	\mathcal{L}_{22}^T & \mathcal{L}_{21}^T & 0 \\
		0 & \mathcal{L}_{32}^T & 	\mathcal{L}_{33}^T & 0 \\
		\mathcal{L}_{41}^T & 0 & 0 & 	\mathcal{L}_{44}^T
	\end{pmatrix},\label{Lo}
\end{equation}
where
\begin{align}\nonumber
	\mathcal{L}_{11}^\omega &=\frac{\eta ^2 (\partial_\omega y-\partial_\omega x)+x(x+y)\,\partial_\omega x  -2\eta  y\, \partial_\omega \eta }{(x+y) \left(x y-\eta ^2\right)} , \\\nonumber
	\mathcal{L}_{22}^\omega &=	\mathcal{L}_{3,3}^\omega
	=\frac{ z\,\partial_\omega z-\delta \,\partial_\omega \delta}{z^2-\delta ^2}, \\
	\mathcal{L}_{14}^\omega &=	\mathcal{L}_{41}^\omega=\frac{ \eta  y\,\partial_\omega x+ \eta  x \,\partial_\omega y-2  x y\,\partial_\omega \eta}{(x+y) \left(\eta ^2-x y\right)} ,\\\nonumber
	\mathcal{L}_{23}^\omega &=	\mathcal{L}_{32}^\omega= 	\frac{ z\,\partial_\omega \delta-\delta \, \partial_\omega z}{z^2-\delta ^2}\\\nonumber
	\mathcal{L}_{44}^\omega &=\frac{\eta ^2
		(\partial_\omega x-\partial_\omega y)+ x (x+y)\,\partial_\omega y-2  \eta  x\,\partial_\omega \eta}{(x+y) \left(x y-\eta ^2\right)}\nonumber
\end{align}
Finally, when the quantum Cramér--Rao bound is saturated, the minimum
achievable variances for the simultaneous estimation of the parameters
$\omega$ and $T$ are given by

\begin{equation}
	\mathrm{Var}(\omega)_{\min}=\frac{\mathcal{F}_{TT}}{\det(\mathcal{F})},
	\qquad
	\mathrm{Var}(T)_{\min}=\frac{\mathcal{F}_{\omega \omega}}{\det(\mathcal{F})} .
\end{equation}
\begin{figure}[H]
	\centering
	\includegraphics[scale=0.8]{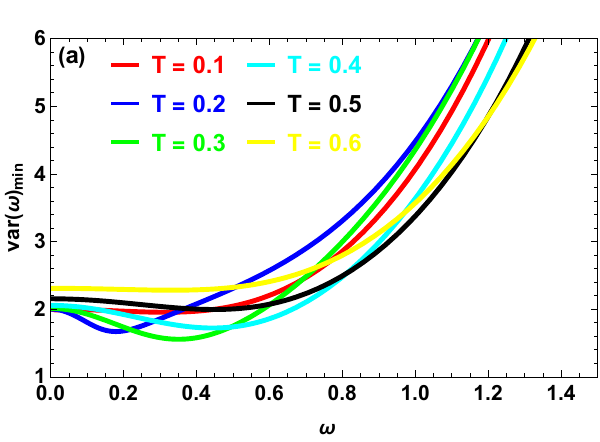}
	\hspace*{0.3cm}
	\includegraphics[scale=0.8]{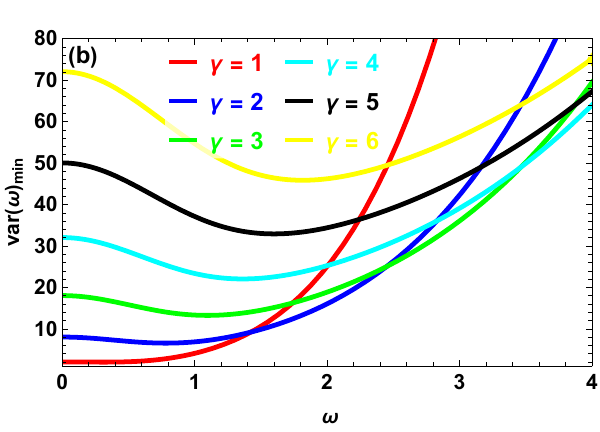}
	\caption{Minimal variance for the simultaneous estimation of $\omega$ as a function of $\omega$: 
		(a) for different values of $T$ with $\gamma = 1$; 
		(b) for different values of $\gamma$ with $T = 0.1$.}
	\label{fig:11}
\end{figure}
The minimal variance associated with the simultaneous estimation of the parameter $\omega$ as a function of $\omega$ for different values of the temperature $T$ with $\gamma=1$ is presented in Fig.~\ref{fig:11}(a). The curves exhibit a non–monotonic behavior, characterized by the presence of a minimum that identifies an optimal estimation region. For small values of $\omega$, the variance slightly decreases before reaching its minimum, indicating an enhancement of the estimation precision. As $\omega$ increases further, the variance grows rapidly due to the reduced sensitivity of the quantum state to variations of the energy splitting parameter. Moreover, increasing the temperature slightly modifies the position and the depth of the minimum, revealing the influence of thermal effects on the estimation accuracy. 

The dependence of $\mathrm{Var}(\omega)_{\min}$ on $\omega$ for several values of the coupling parameter $\gamma$ at fixed $T=0.1$ is illustrated in Fig.~\ref{fig:11}(b). A pronounced minimum is observed for each curve, indicating the existence of an optimal value of $\omega$ that maximizes the estimation precision. Increasing $\gamma$ significantly enlarges the magnitude of the variance and shifts the optimal region, reflecting the strong impact of the gravitational coupling on the structure of the energy spectrum and consequently on the attainable precision in the simultaneous estimation protocol. 
\begin{figure}[H]
	\centering
	\includegraphics[scale=0.8]{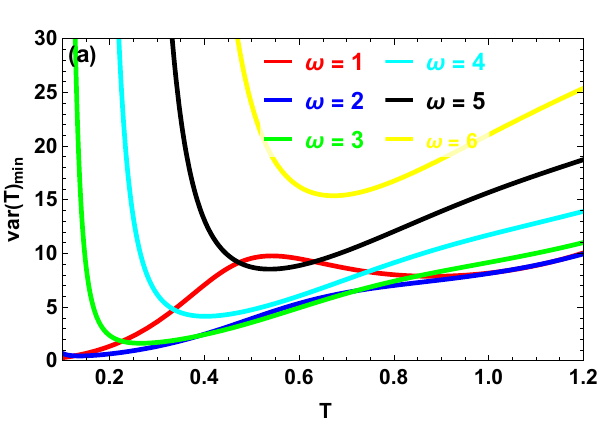}
	\hspace*{0.3cm}
	\includegraphics[scale=0.8]{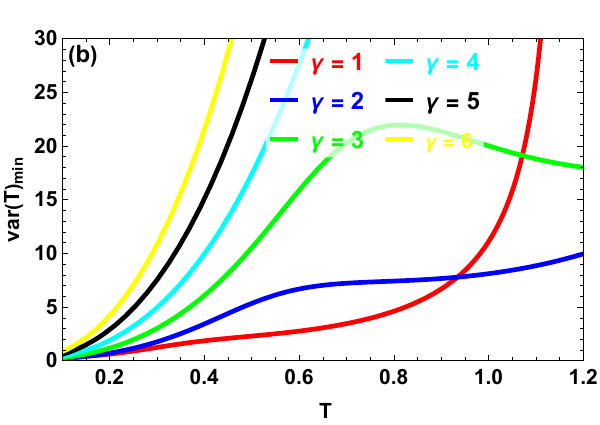}
	\caption{Minimal variance for the simultaneous estimation of $T$ as a function of $T$: 
		(a) for different values of $\omega$ with $\gamma = 2$; 
		(b) for different values of $\gamma$ with $\omega = 1.5$.}
	\label{fig:12}
\end{figure}
The behavior of the minimal variance associated with the simultaneous estimation of the temperature $T$ as a function of $T$ for different values of the energy splitting parameter $\omega$ with $\gamma=2$ is illustrated in Fig.~\ref{fig:12}(a). The curves display a non–monotonic profile characterized by the presence of a minimum, which identifies an optimal temperature region where the estimation precision is enhanced. For small temperatures, the variance remains relatively low, indicating a high sensitivity of the quantum state to thermal fluctuations. As $T$ increases, the variance grows progressively due to the redistribution of the thermal populations among the energy levels. Moreover, increasing $\omega$ modifies both the position and the depth of the optimal region, reflecting the influence of the system energy gap on the estimation precision. 

The dependence of $\mathrm{Var}(T)_{\min}$ on $T$ for several values of the coupling parameter $\gamma$ at fixed $\omega=1.5$ is presented in Fig.~\ref{fig:12}(b). The results indicate that the interaction strength significantly affects the estimation accuracy. In particular, larger values of $\gamma$ tend to increase the variance and shift the optimal estimation region, highlighting the important role played by the gravitational coupling in determining the thermal sensitivity of the system.

Furthermore, when the parameters $\omega$ and $T$ are estimated individually, the Cramér–Rao bound takes the form
\begin{equation}
	\mathrm{Var}(\omega)^{\mathrm{Ind}} \geq \mathcal{F}^{-1}_{\omega\omega}, 
	\qquad
	\mathrm{Var}(T)_{\mathrm{Ind}} \geq \mathcal{F}^{-1}_{TT},
\end{equation}
and the corresponding minimal variances are given by
\begin{equation}
	\mathrm{Var}(\omega)^{\mathrm{Ind}}_{\min} = \mathcal{F}^{-1}_{\omega\omega}, 
	\qquad
	\mathrm{Var}(T)^{\mathrm{Ind}}_{\min} = \mathcal{F}^{-1}_{TT}.
\end{equation} 
\begin{figure}[H]
	\centering
	\includegraphics[scale=0.8]{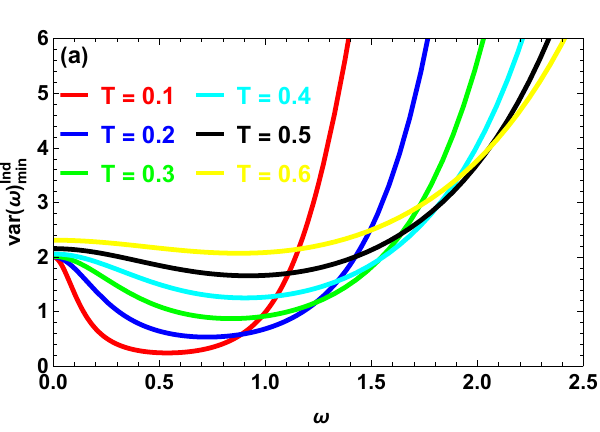}
	\hspace*{0.3cm}
	\includegraphics[scale=0.8]{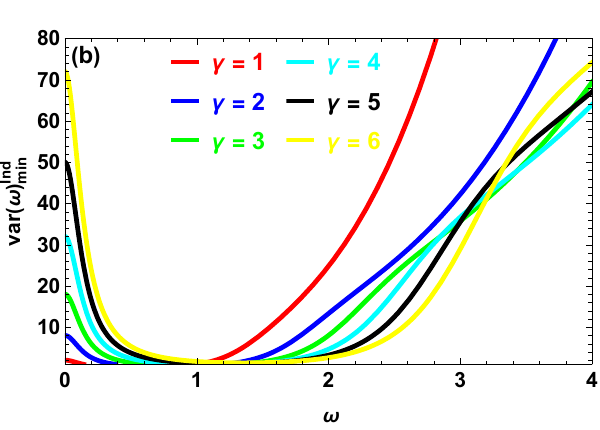}
	\caption{Minimal variance for the  individual estimation of $\omega$ as a function of $\omega$: 
		(a) for different values of $T$ with $\gamma = 1$; 
		(b) for different values of $\omega$ with $T= 0.1$.}
	\label{fig:13}
\end{figure}
Fig.~\ref{fig:13}(a) displays the minimal variance associated with the individual estimation of the parameter $\omega$ as a function of $\omega$ for several values of the temperature $T$ with $\gamma=1$. The curves exhibit a non–monotonic behavior characterized by the presence of a minimum, which identifies an optimal region where the estimation precision is maximized. For small values of $\omega$, the variance decreases and reaches its minimum before increasing again as $\omega$ grows. This behavior reflects the variation of the sensitivity of the quantum state to changes in the energy splitting parameter. In addition, increasing the temperature modifies both the depth and the position of the optimal region, indicating the influence of thermal fluctuations on the estimation precision. 

Fig.~\ref{fig:13}(b) illustrates the dependence of $\mathrm{Var}(\omega_{\mathrm{Ind}})_{\min}$ on $\omega$ for different values of the coupling parameter $\gamma$ at fixed $T=0.1$. A clear minimum appears for each curve, indicating the existence of an optimal value of $\omega$ that minimizes the estimation error. Increasing $\gamma$ significantly affects the magnitude of the variance and shifts the optimal region, highlighting the important role played by the interaction strength in determining the achievable precision in the individual estimation strategy. 
\begin{figure}[H]
	\centering
	\includegraphics[scale=0.8]{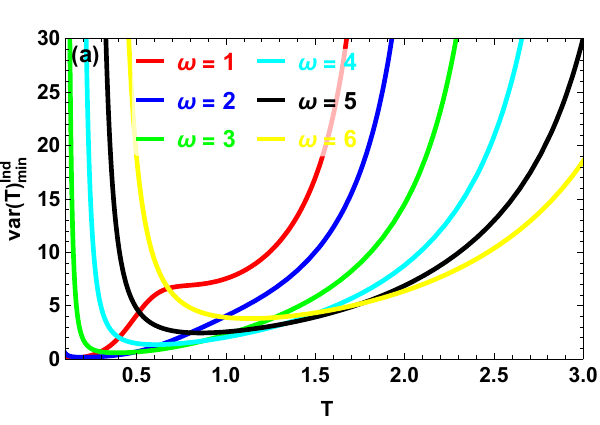}
	\hspace*{0.3cm}
	\includegraphics[scale=0.8]{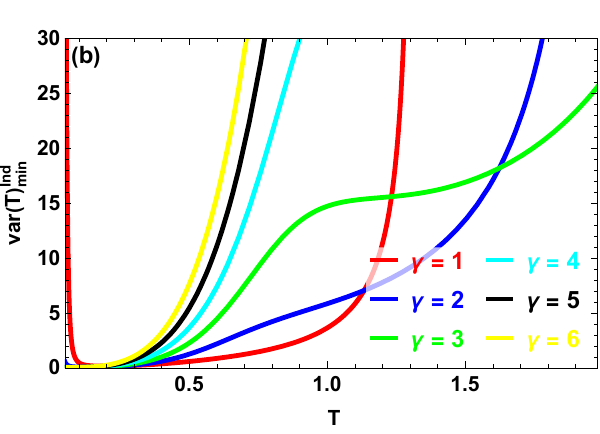}
	\caption{Minimal variance for the  individual estimation of $T$ as a function of $T$: 
		(a) for different values of $\omega$ with $\gamma =2$; 
		(b) for different values of $\gamma$ with $\omega= 1.5$.}
	\label{fig:14}
\end{figure}
Fig.~\ref{fig:14}(a) highlights how the minimal variance associated with the individual estimation of the temperature $T$ evolves with $T$ for several values of the energy splitting parameter $\omega$ at fixed $\gamma=2$. A clear dependence on the value of $\omega$ is observed: larger energy gaps tend to shift the region of low variance toward higher temperatures while simultaneously increasing the overall magnitude of the estimation error. This behavior reflects the modification of the thermal population distribution induced by the energy spectrum of the system, which directly affects the sensitivity of the state to temperature variations. 

A different behavior emerges when varying the coupling parameter $\gamma$, as shown in Fig.~\ref{fig:14}(b) for $\omega=1.5$. The curves indicate that the interaction strength plays a crucial role in shaping the temperature dependence of the estimation precision. In particular, stronger coupling generally leads to a faster growth of the variance with increasing temperature, revealing that the interaction tends to reduce the thermal sensitivity of the system in the individual estimation protocol. 

The parameters $\omega$ and $T$, by analyzing the variance ratio between the two estimation methods.
\begin{equation}
	\Gamma(\omega,T)=\frac{\Delta(\omega,T)_{\mathrm{Sim}}}{\Delta(\omega,T)_{\mathrm{Ind}}}
	=\frac
	{\frac{1}{2}\left(\mathrm{Var}(\omega)_{\min}+\mathrm{Var}(T)_{\min}\right)}{\left(\mathrm{Var}(\omega)_{\min}^{\mathrm{Ind}}+\mathrm{Var}(T)_{\min}^{\mathrm{Ind}}\right)} .
\end{equation}

\begin{figure}[H]
	\centering
	\includegraphics[scale=0.8]{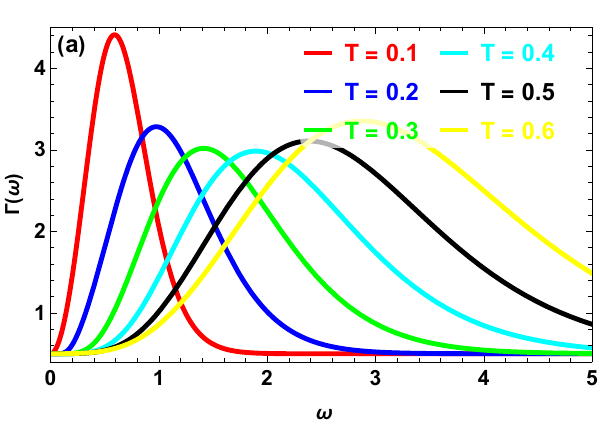}
	\hspace*{0.3cm}
	\includegraphics[scale=0.8]{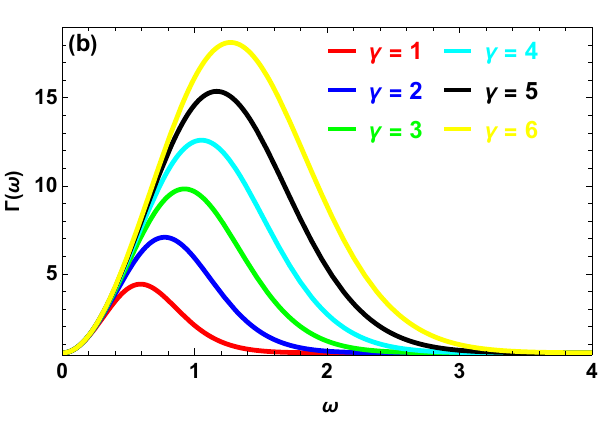}
	\caption{The ratio between the minimal total variances in the estimation of the parameters $\omega$ and $T$ as a function of $\omega$: 
		(a) for different values of $\omega$ with $\gamma = 1$; 
		(b) for different values of $\gamma$ with $T=0.1$.}
	\label{fig:15}
\end{figure}
Fig.~\ref{fig:15}(a) compares the performance of the simultaneous and individual estimation schemes through the ratio $\Gamma(\omega)$ as a function of $\omega$ for several values of the temperature $T$ with $\gamma=1$. The curves exhibit a pronounced peak whose position shifts toward larger values of $\omega$ as the temperature increases. In the regions where $\Gamma(\omega)>1$, the individual estimation protocol provides a higher precision than the simultaneous strategy, while $\Gamma(\omega)<1$ indicates the superiority of the simultaneous estimation approach. The results show that the relative efficiency of the two schemes strongly depends on both the energy splitting parameter and the thermal environment. 

The effect of the coupling parameter $\gamma$ on the ratio $\Gamma(\omega)$ at fixed temperature $T=0.1$ is illustrated in Fig.~\ref{fig:15}(b). In this case, the curves reveal a rapid increase of the ratio with $\omega$, indicating that the advantage of the simultaneous estimation scheme is mainly restricted to the low–$\omega$ region. As the interaction strength grows, the crossover between the two estimation strategies occurs at larger values of $\omega$, highlighting the role of the gravitational coupling in determining the optimal estimation protocol. 
\begin{figure}[H]
	\centering
	\includegraphics[scale=0.8]{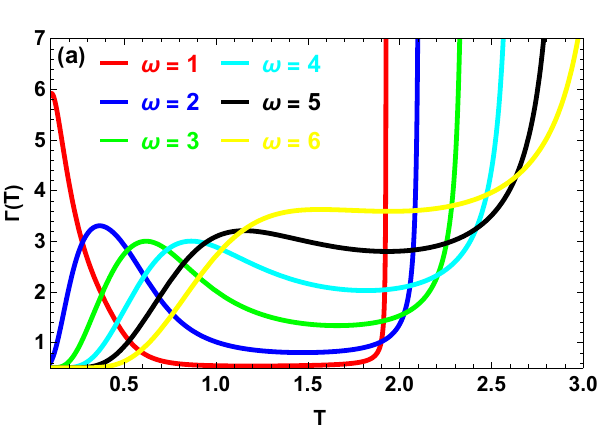}
	\hspace*{0.3cm}
	\includegraphics[scale=0.8]{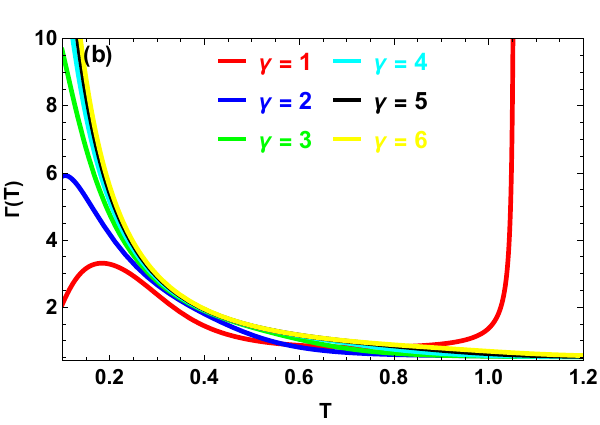}
	\caption{The ratio between the minimal total variances in the estimation of the parameters $\omega$ and $T$ as a function of $T$: 
		(a) for different values of $\omega$ with $\gamma = 2$; 
		(b) for different values of $\gamma$ with $\omega=1$.}
	\label{fig:16}
\end{figure}
In Fig.~\ref{fig:16}(a), the ratio $\Gamma(T)$ is plotted as a function of the temperature for several values of the energy splitting parameter $\omega$ with $\gamma=2$. The curves reveal a strong dependence of the relative performance of the estimation strategies on the thermal regime. In particular, the ratio may exceed unity at low temperatures for certain values of $\omega$, indicating that the individual estimation can provide better precision in this region. As the temperature increases, $\Gamma(T)$ generally decreases and approaches values below one, showing that the simultaneous estimation scheme becomes progressively more advantageous.

In Fig.~\ref{fig:16}(b), the influence of the coupling parameter $\gamma$ on $\Gamma(T)$ is analyzed at fixed $\omega=1$. A rapid decrease of the ratio with increasing temperature is observed for all interaction strengths. This behavior indicates that thermal fluctuations tend to enhance the advantage of the simultaneous estimation strategy. Furthermore, larger values of $\gamma$ slightly modify the temperature interval where the transition between the two estimation schemes occurs, highlighting the role of the interaction strength in determining the optimal estimation protocol.
\subsection{Estimation of parameters $\omega$ and $\gamma$}
In this subsection, we investigate the simultaneous estimation of the two parameters 
$\Omega$ and $\gamma$. The corresponding Quantum Fisher Information Matrix (QFIM) 
can be written as
\begin{equation}
	\mathcal{F}=
	\begin{pmatrix}
	\mathcal{F}_{\omega\omega} & \mathcal{F}_{\omega\gamma}\\
		\mathcal{F}_{\gamma\omega} & \mathcal{F}_{\gamma\gamma}
	\end{pmatrix}
	=
	\begin{pmatrix}
		2\,\mathrm{vec}[\partial_{\omega}\varrho]^{\dagger T}\,\mathfrak{R}^{-1}\,
		\mathrm{vec}[\partial_{\omega}\varrho] &
		2\,\mathrm{vec}[\partial_{\omega}\varrho]^{\dagger T}\,\mathfrak{R}^{-1}\,
		\mathrm{vec}[\partial_{\gamma}\varrho] \\
		2\,\mathrm{vec}[\partial_{\gamma}\varrho]^{\dagger T}\,\mathfrak{R}^{-1}\,
		\mathrm{vec}[\partial_{\omega}\varrho] &
		2\,\mathrm{vec}[\partial_{\gamma}\varrho]^{\dagger T}\,\mathfrak{R}^{-1}\,
		\mathrm{vec}[\partial_{\gamma}\varrho]
	\end{pmatrix},
\end{equation}
where $\mathrm{vec}[\partial_{\omega}\rho]$ and 
$\mathrm{vec}[\partial_{\gamma}\rho]$ are defined in Eqs.~(\ref{veco}) and (\ref{vecg}), respectively. The elements of the matrix $\mathcal{F}$ are presented in Appendix~\ref{Appendix A}. 
According to the multiparameter quantum Cramér--Rao bound, the variances satisfy
\begin{equation}
	\mathrm{Var}(\omega) \geq \frac{\mathcal{F}_{\gamma\gamma}}{\det(\mathcal{F})},
	\qquad
	\mathrm{Var}(\gamma) \geq \frac{\mathcal{F}_{\omega\omega}}{\det(\mathcal{F})},
\end{equation}
and
\begin{equation}
	\left(\mathrm{Var}(\omega)-\frac{\mathcal{F}_{\gamma\gamma}}{\det(\mathcal{F})}\right)
	\left(\mathrm{Var}(\gamma)-\frac{\mathcal{F}_{\omega\omega}}{\det(\mathcal{F})}\right)
	\geq
	\left(\mathrm{Cov}(\omega,\gamma)+\frac{\mathcal{F}_{\omega\gamma}}{\det(\mathcal{F})}\right)^2 .
\end{equation}

The symmetric logarithmic derivative (SLD) operators $\mathcal{L}_{\omega}$ and 
$\mathcal{L}_{\gamma}$ are given in Eqs.~(\ref{Lo}) and (\ref{Lg}). Consequently, the minimum 
achievable variances are
\begin{equation}
	\mathrm{Var}(\omega)_{\min}=\frac{\mathcal{F}_{\gamma\gamma}}{\det(\mathcal{F})},
	\qquad
	\mathrm{Var}(\gamma)_{\min}=\frac{\mathcal{F}_{\omega\omega}}{\det(\mathcal{F})} .
\end{equation}
\begin{figure}[H]
	\centering
	\includegraphics[scale=0.8]{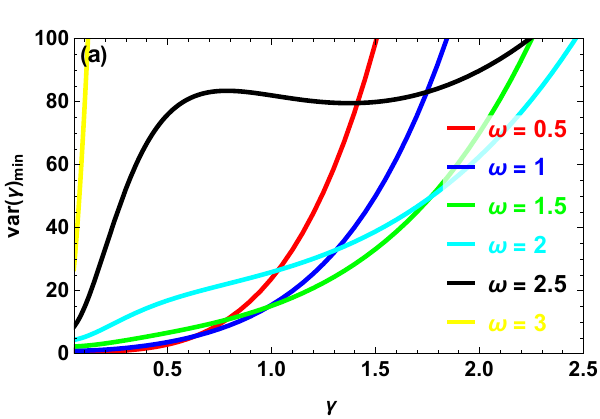}
	\hspace*{0.3cm}
	\includegraphics[scale=0.8]{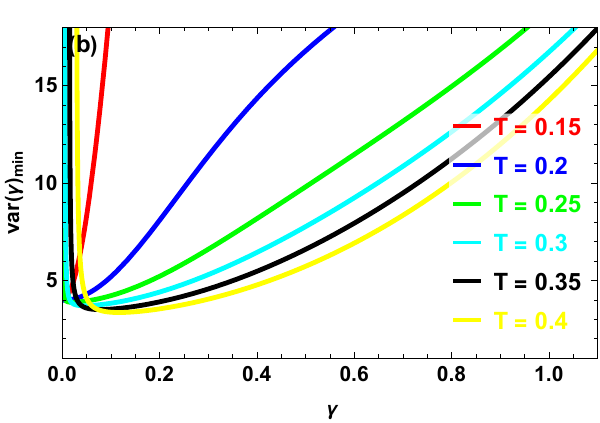}
	\caption{Minimal variance for the simultaneous estimation of $\gamma$ as a function of $\gamma$: 
		(a) for different values of $\omega$ with $T=0.2$; 
		(b) for different values of $\gamma$ with $\omega = 2$.}
	\label{fig:17}
\end{figure}
In Fig.~\ref{fig:17}(a), the minimal variance associated with the simultaneous estimation of the coupling parameter $\gamma$ is shown as a function of $\gamma$ for several values of the energy splitting parameter $\omega$ at fixed temperature $T=0.2$. The curves reveal that the estimation precision is strongly influenced by the value of $\omega$. For small interaction strengths, the variance remains relatively low, indicating a favorable estimation region. However, as $\gamma$ increases, the variance grows significantly, reflecting the reduced sensitivity of the quantum state to variations of the coupling parameter. In addition, larger values of $\omega$ tend to increase the overall magnitude of the variance, highlighting the impact of the energy gap on the achievable estimation precision. 

In Fig.~\ref{fig:17}(b), the dependence of the minimal variance $\mathrm{Var}(\gamma)_{\min}$ on $\gamma$ is examined for different temperatures with $\omega=2$. The curves show that increasing the temperature generally leads to larger variances, indicating a degradation of the estimation precision due to thermal fluctuations. Nevertheless, for moderate values of $\gamma$, the variance remains relatively small, suggesting that this region provides more favorable conditions for the simultaneous estimation of the parameters $\omega$ and $\gamma$.
\begin{figure}[H]
	\centering
	\includegraphics[scale=0.8]{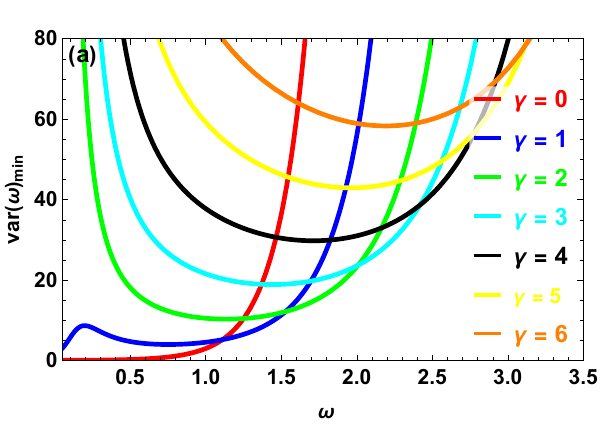}
	\hspace*{0.3cm}
	\includegraphics[scale=0.8]{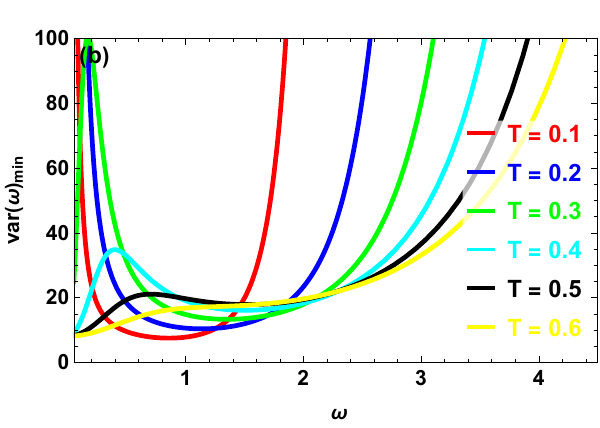}
	\caption{Minimal variance for the simultaneous estimation of $\omega$ as a function of $\omega$: 
		(a) for different values of $\gamma$ with $T=0.2$; 
		(b) for different values of $T$ with $\gamma = 2$.}
	\label{fig:18}
\end{figure}
The minimal variance associated with the simultaneous estimation of the parameter $\omega$ as a function of $\omega$ for several values of the coupling parameter $\gamma$ at fixed temperature $T=0.2$ is presented in Fig.~\ref{fig:18}(a). The curves exhibit a non–monotonic behavior characterized by the presence of a minimum, indicating the existence of an optimal region where the estimation precision is enhanced. For small values of $\omega$, the variance remains relatively low, whereas it increases rapidly for larger $\omega$, reflecting the reduced sensitivity of the quantum state to variations of the energy splitting parameter. Moreover, increasing the coupling parameter $\gamma$ significantly modifies both the magnitude and the position of the minimum, highlighting the important role played by the interaction strength in the estimation process. 

The dependence of the minimal variance $\mathrm{Var}(\omega)_{\min}$ on $\omega$ for different values of the temperature $T$ with $\gamma=2$ is shown in Fig.~\ref{fig:18}(b). Thermal effects strongly influence the estimation precision: increasing the temperature generally leads to larger variances and shifts the optimal estimation region. This behavior originates from the redistribution of the thermal populations among the energy levels, which affects the sensitivity of the system to variations of the parameter $\omega$. 

We analyze the minimum variances associated with the individual estimation of the parameters $\gamma$ and $\omega$, denoted by 
$\mathrm{Var}(\gamma)^{\mathrm{Ind}}_{\min}$ and 
$\mathrm{Var}(\omega)^{\mathrm{Ind}}_{\min}$, respectively. These quantities are given by
\begin{equation}
	\mathrm{Var}(\omega)_{\mathrm{Ind}}^{\min}=\mathcal{F}^{-1}_{\omega\omega},
	\qquad
	\mathrm{Var}(\gamma)_{\mathrm{Ind}}^{\min}=\mathcal{F}^{-1}_{\gamma\gamma}.
\end{equation}
\begin{figure}[H]
	\centering
	\includegraphics[scale=0.8]{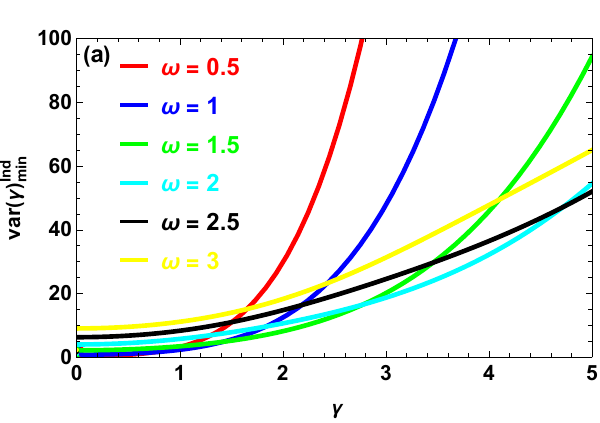}
	\hspace*{0.3cm}
	\includegraphics[scale=0.8]{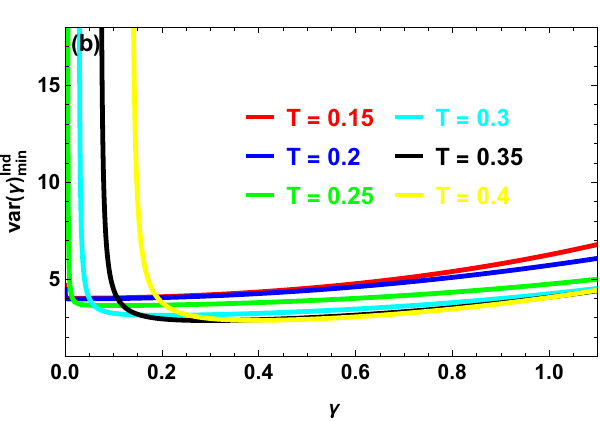}
	\caption{Minimal variance for the  individual estimation of $\gamma$ as a function of $\gamma$: 
		(a) for different values of $\omega$ with $T=0.2$; 
		(b) for different values of $T$ with $\omega=2$.}
	\label{fig:19}
\end{figure}
The individual estimation precision of the coupling parameter $\gamma$ as a function of $\gamma$ for several values of the energy splitting parameter $\omega$ at fixed temperature $T=0.2$ is illustrated in Fig.~\ref{fig:19}(a). The curves show that the estimation accuracy strongly depends on the value of $\omega$. For weak interaction strengths, the variance remains relatively small, indicating a favorable estimation region. However, as $\gamma$ increases, the variance rises significantly, revealing that the sensitivity of the quantum state to variations of the coupling parameter gradually decreases. Larger values of $\omega$ also lead to an overall increase in the variance, emphasizing the influence of the energy gap on the attainable precision. 

The influence of the temperature on the individual estimation of $\gamma$ with $\omega=2$ is presented in Fig.~\ref{fig:19}(b). The results indicate that thermal effects play a significant role in determining the estimation precision. In particular, increasing the temperature generally leads to larger variances for most values of $\gamma$, reflecting the detrimental impact of thermal fluctuations. Nevertheless, the variance remains relatively small in the region of moderate interaction strengths, suggesting that this interval provides more suitable conditions for parameter estimation. 
\begin{figure}[H]
	\centering
	\includegraphics[scale=0.8]{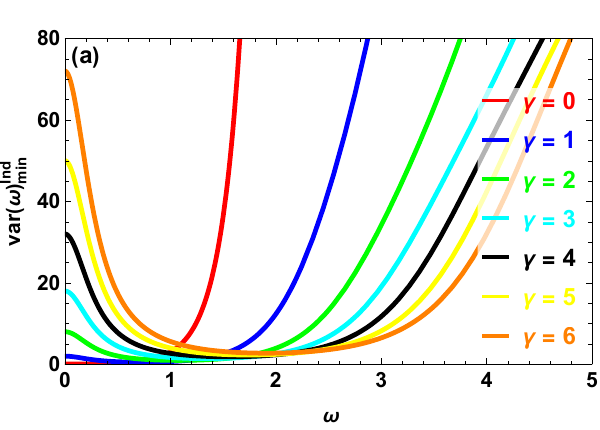}
	\hspace*{0.3cm}
	\includegraphics[scale=0.8]{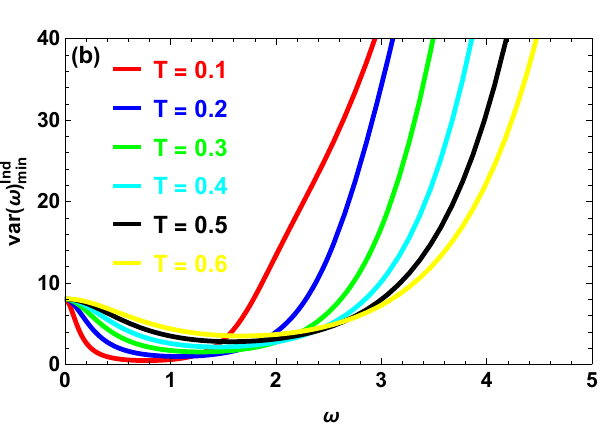}
	\caption{Minimal variance for the  individual estimation of $\omega$ as a function of $\omega$: 
		(a) for different values of $\gamma$ with $T=0.2$; 
		(b) for different values of $T$ with $\omega=2$.}
	\label{fig:20}
\end{figure}
The behavior of the minimal variance associated with the individual estimation of the parameter $\omega$ for several values of the coupling parameter $\gamma$ at fixed temperature $T=0.2$ is reported in Fig.~\ref{fig:20}(a). A pronounced minimum appears for each curve, indicating the presence of an optimal region of $\omega$ where the estimation precision is improved. For small values of $\omega$, the variance decreases rapidly before reaching this optimal point, after which it increases again for larger $\omega$. The position of this minimum slightly shifts as the interaction strength varies, showing that the coupling parameter plays a relevant role in determining the sensitivity of the system to changes in $\omega$. 

The influence of temperature on the individual estimation of $\omega$ with $\gamma=2$ is displayed in Fig.~\ref{fig:20}(b). The results show that thermal effects modify the curvature of the variance profiles while preserving the existence of an optimal estimation region. In particular, increasing the temperature tends to enlarge the minimal variance and slightly shift the optimal value of $\omega$. This behavior can be attributed to the redistribution of thermal populations among the system energy levels, which directly affects the sensitivity of the quantum state to variations of the parameter $\omega$. 

The ratio $R(\omega,\gamma)$ is defined as
\begin{equation}
	\Gamma(\omega,\gamma)=\frac{\Delta(\omega,\gamma)_{\mathrm{Sim}}}{\Delta(\omega,\gamma)_{\mathrm{Ind}}}
	=\frac
	{\frac{1}{2}\left(\mathrm{Var}(\omega)_{\min}+\mathrm{Var}(\gamma)_{\min}\right)}{\left(\mathrm{Var}(\omega)_{\min}^{\mathrm{Ind}}+\mathrm{Var}(\gamma)_{\min}^{\mathrm{Ind}}\right)} .
\end{equation}

The ratio $\Gamma(\omega,\gamma)$, illustrated in Fig.~15, allows us to compare the minimum total variances obtained from the simultaneous estimation strategy with those resulting from individual estimations of the parameters $\omega$ and $\gamma$. 
\begin{figure}[H]
	\centering
	\includegraphics[scale=0.8]{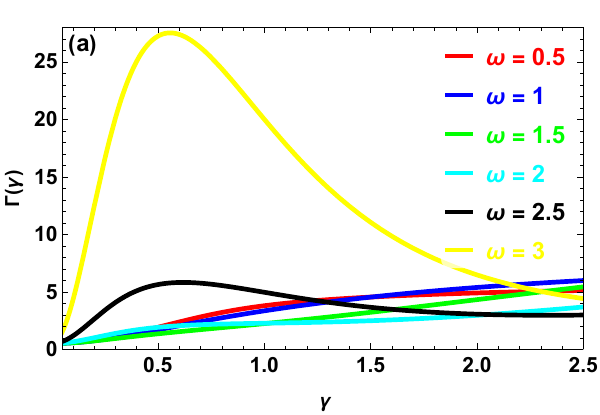}
	\hspace*{0.3cm}
	\includegraphics[scale=0.8]{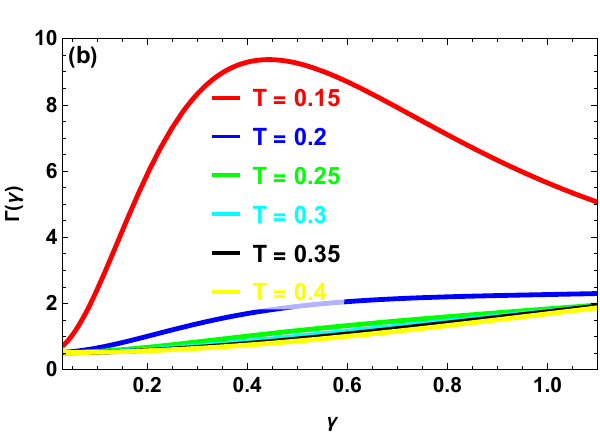}
	\caption{The ratio between the minimal total variances in the estimation of the parameters $\omega$ and $\gamma$ as a function of $\gamma$: 
		(a) for different values of $\omega$ with $T = 0.2$; 
		(b) for different values of $T$ with $\omega=2$.}
	\label{fig:21}
\end{figure}
The relative efficiency between the simultaneous and individual estimation strategies for the parameters $\omega$ and $\gamma$ is analyzed through the ratio $\Gamma(\gamma)$ shown in Fig.~\ref{fig:21}(a). For several values of the energy splitting parameter $\omega$ at fixed temperature $T=0.2$, the curves reveal a pronounced peak structure indicating a strong variation in the relative performance of the two estimation schemes. In particular, the ratio increases significantly in an intermediate interaction regime, especially for larger values of $\omega$, suggesting that the advantage of one strategy over the other strongly depends on the interplay between the coupling strength and the energy gap of the system. 

The effect of temperature on this ratio is illustrated in Fig.~\ref{fig:21}(b) for $\omega=2$. The curves show that increasing the temperature generally reduces the magnitude of $\Gamma(\gamma)$ and smooths its variation with the interaction strength. This behavior indicates that thermal fluctuations tend to diminish the difference between the simultaneous and individual estimation schemes, leading to a more uniform estimation performance over a wide range of $\gamma$. 
\begin{figure}[H]
	\centering
	\includegraphics[scale=0.8]{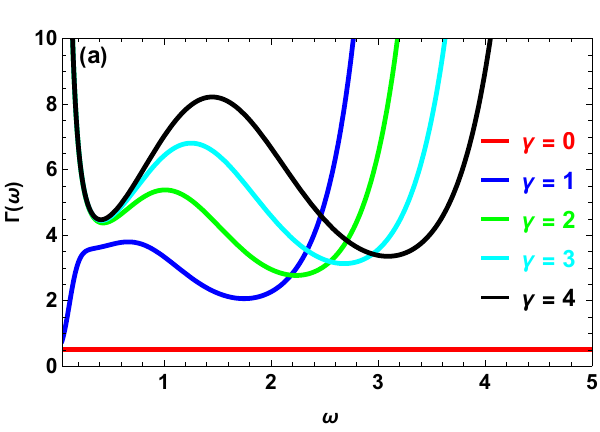}
	\hspace*{0.3cm}
	\includegraphics[scale=0.8]{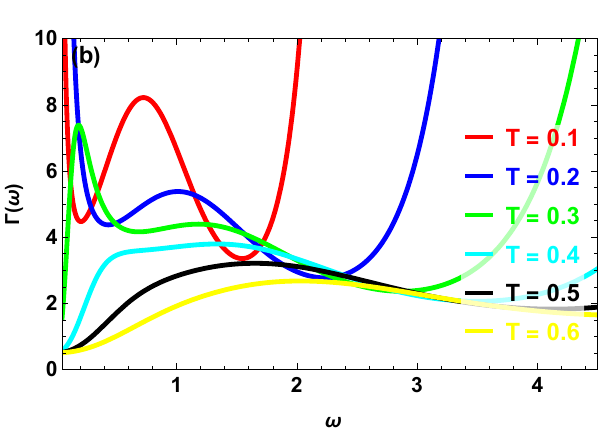}
	\caption{The ratio between the minimal total variances in the estimation of the parameters $\omega$ and $\gamma$ as a function of $\omega$: 
		(a) for different values of $\gamma$ with $T = 0.2$; 
		(b) for different values of $T$ with $\omega=2$.}
	\label{fig:22}
\end{figure}
Fig.~\ref{fig:22}(a) illustrates the behavior of the ratio $\Gamma(\omega)$ as a function of $\omega$ for several values of the coupling parameter $\gamma$ at fixed temperature $T=0.2$. The curves display a pronounced non–monotonic structure, revealing that the relative efficiency between the simultaneous and individual estimation strategies strongly depends on the interaction strength. In particular, the position and amplitude of the maxima shift with increasing $\gamma$, indicating that the coupling parameter significantly influences the optimal estimation regime for the parameter $\omega$. 

Fig.~\ref{fig:22}(b) shows the corresponding dependence of $\Gamma(\omega)$ for different temperatures with $\omega=2$. Thermal effects lead to noticeable modifications in the shape of the curves, especially at low temperatures where the ratio can reach relatively large values. As the temperature increases, the variation of $\Gamma(\omega)$ becomes smoother, suggesting that thermal fluctuations tend to reduce the contrast between the simultaneous and individual estimation strategies. 

In this section, we have analyzed the quantum estimation of the three fundamental parameters of the gravcat model, namely the gravitational coupling $\gamma$, the energy splitting $\omega$, and the temperature $T$. Using the quantum Fisher information matrix formalism, both individual and simultaneous estimation strategies were investigated for different pairs of parameters.
The results obtained in the first subsection show that the joint estimation of $(\gamma,T)$ is strongly affected by the interaction strength and the thermal environment, with specific regions where the estimation precision is significantly enhanced. In the second subsection, the estimation of $(\omega,T)$ reveals that the energy gap plays a crucial role in determining the thermal sensitivity of the system. The third subsection, devoted to the estimation of $(\gamma,\omega)$, highlights the interplay between the coupling strength and the energy splitting in shaping the attainable precision.
A systematic comparison between simultaneous and individual estimation schemes shows that neither strategy is universally optimal. Instead, their relative performance depends on the physical regime defined by the values of $\gamma$, $\omega$, and $T$. These results provide a comprehensive picture of the metrological capabilities of the gravcat model and clarify how the different physical parameters influence the achievable estimation precision.
\section{Quantum thermodynamics}\label{sec5}
In the previous section, we analyzed the quantum estimation of the fundamental parameters of the gravcat model, namely the gravitational coupling $\gamma$, the energy splitting $\omega$, and the temperature $T$. By employing the quantum Fisher information matrix formalism, the optimal regions for the estimation of these parameters were identified under different physical conditions.
In this section, we investigate the thermodynamic behavior of the gravcat system by considering a quantum Stirling cycle. In particular, we explore how the thermodynamic performance of the cycle is influenced by the system parameters previously estimated. The optimal values of $\gamma$, $\omega$, and $T$ obtained from the quantum metrological analysis are used to characterize the working conditions of the cycle.
The Stirling cycle, consisting of two isothermal and two isochoric processes, provides a suitable framework for studying the interplay between quantum thermodynamics and parameter estimation. By analyzing quantities such as the exchanged heat, the work performed during the cycle, and the corresponding efficiency, we aim to understand how the optimal estimation regimes identified earlier affect the thermodynamic properties of the gravcat system.

Before introducing the thermodynamic quantities characterizing the system, it is useful to determine the occupation probabilities of the energy levels. For a quantum system in thermal equilibrium with a heat bath at temperature $T$, the occupation probabilities $P_n$ of the energy eigenstates are given by the Boltzmann distribution

\begin{equation}
	P_n=\dfrac{e^{-\beta E_n}}{\sum_{n=1}^{4} e^{-\beta E_n}}=\dfrac{e^{-\beta E_n}}{Z},\label{prob}
\end{equation}
These probabilities play a central role in the thermodynamic description of the system since they allow the calculation of the relevant thermodynamic quantities. In particular, the internal energy and the von Neumann entropy can be expressed in terms of the occupation probabilities of the system.
Using Eq.~(\ref{vap}), we obtain the following relations
\begin{align*}
	P_1&=\frac{e^{\,-\beta\,E_1}}{Z},\qquad
	P_2=\frac{e^{\,-\beta\,E_2}}{Z}\\
		P_3&=\frac{e^{\,-\beta\,E_3}}{Z},\,\qquad
		P_4=\frac{e^{\,-\beta\,E_4}}{Z},
\end{align*}
The von Neumann entropy of the system is defined as 
\begin{equation}
	S = - \mathrm{Tr}\left( \varrho^T \ln \varrho^T\right)  = \sum_{n=1}^{4} P_n \ln P_n .
	\label{eq:entropy}
\end{equation}
Here, $P_n$ denotes the occupation probability introduced in Eq.~(\ref{prob}). 
In this work, we assume that the system is in thermal equilibrium. 
Under this condition, the state of the system can be described by the following density matrix in Eq.~(\ref{rho}).
The system’s internal energy is given by
\begin{equation}
	U = \mathrm{Tr}\left( \varrho^T \mathcal{H}\right)  = \sum_{n=1}^{4} P_n E_n
	= P_1E_1 + P_2E_2 + P_3E_3 + P_4E_4 .
\end{equation}
\subsection{Quantum stirling cycle}

\begin{figure}[H]
	\centering
	\includegraphics[width=0.5\linewidth]{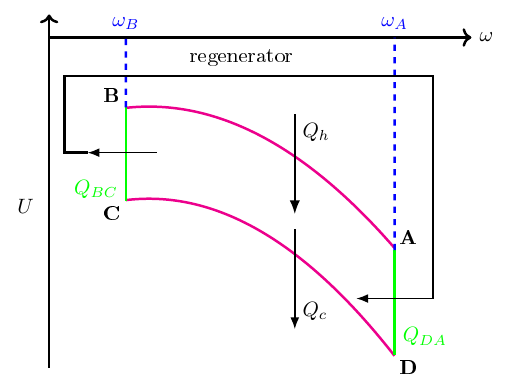}
	\caption{The diagram of the Stirling cycle \cite{huang2014effects}, where $Q_{h}$ denotes the absorbed heat, $Q_{c}$ the released heat, $T_{h}$ the hot-bath temperature, and $T_{c}$ the cold-bath temperature.}
	\label{CS}
\end{figure}

The quantum Stirling cycle (SC) consists of four stages: two quantum isothermal
processes and two isochoric processes (see Fig.~(\ref{CS})). In the isothermal stages,
the magnetic field is varied, while in the isochoric stages the coupling with
the thermal baths is changed. The four steps are described as follows.
\textbf{A $\rightarrow$ B: Isothermal process.}  
The system interacts with a hot heat bath at temperature $T_h$, and the
magnetic field varies from $J_A$ to $J_B$. During this stage, heat is exchanged
between the system and the bath. The absorbed heat is given by
\begin{equation}
	Q_{AB}=T_h\int_A^B dS = T_h (S_B-S_A),
\end{equation}
where the entropy $S$ is obtained from Eq.~(\ref{eq:entropy}).

\textbf{B $\rightarrow$ C: Isochoric process.}  
The magnetic field remains constant at $J_B$, while the temperature decreases
from $T_h$ to $T_c$. The exchanged heat is
\begin{equation}
	Q_{BC}=U_C-U_B = U(J_B,T_c)-U(J_B,T_h).
\end{equation}

\textbf{C $\rightarrow$ D: Isothermal process.}  
The magnetic field varies from $J_B$ to $J_A$ while the system is in contact
with a cold bath at temperature $T_c$. The heat exchanged is
\begin{equation}
	Q_{CD}=T_c\int_C^D dS = T_c (S_D-S_C),
\end{equation}
where the entropy is given by Eq.~(\ref{eq:entropy}).

\textbf{D $\rightarrow$ A: Isochoric process.}  
The magnetic field remains constant at $J_A$, while the temperature increases
from $T_c$ to $T_h$. The absorbed heat is
\begin{equation}
	Q_{DA}=U_A-U_D = U(J_A,T_h)-U(J_A,T_c).
\end{equation}

Over a complete cycle, the total exchanged heat with the hot and cold baths
is given by
\begin{equation}
	Q_h = Q_{AB}+Q_{DA}, \qquad Q_c = Q_{BC}+Q_{CD}.
\end{equation}

The input and output heat are
\begin{equation}
	Q_{AB}=Q_{\text{in}}, \qquad Q_{CD}=Q_{\text{out}} .
\end{equation}

The work performed during one cycle is
\begin{equation}
	W = Q_h + Q_c .
\end{equation}

The efficiency of the quantum heat engine (QHE) is defined as
\begin{equation}
	\eta = \frac{W}{Q_h}.
\end{equation}
\begin{figure}[H]
	\centering
	\includegraphics[scale=0.8]{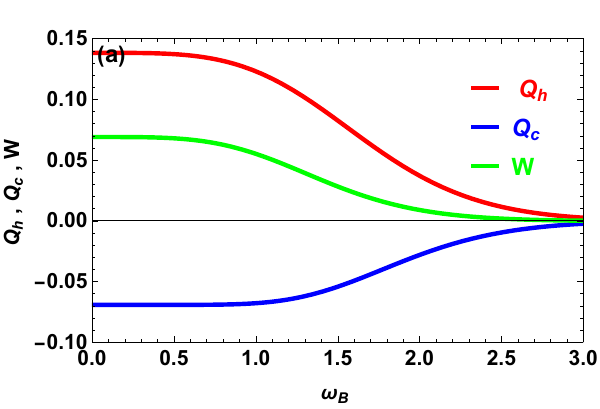}
	\hspace*{0.3cm}
	\includegraphics[scale=0.8]{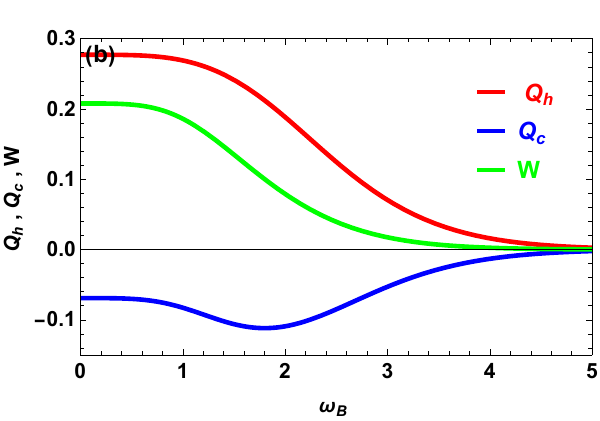}
	\caption{Variation of the heat absorbed $Q_h$, the heat released $Q_c$, and the work done $W$: 
		(a) for $T_h=2T_c$, $\gamma=3$, and $\omega_A=3$; 
		(b) for $T_h=4T_c$, $\gamma=3$, and $\omega_A=5$.}
	\label{fig:7}
\end{figure}
The variation of the absorbed heat $Q_h$, the released heat $Q_c$, and the work $W$ as functions of the parameter $\omega_B$ is illustrated in Fig.~\ref{fig:7} for two different thermal gradients between the hot and cold reservoirs. In both cases, the absorbed heat remains positive while the released heat is negative, confirming that the system operates as a heat engine. The work produced during the cycle is also positive over a certain range of $\omega_B$, indicating that useful work is extracted from the thermal reservoirs.

A clear influence of the temperature difference between the reservoirs can be observed when comparing the two panels. For the case $T_h=2T_c$ (Fig.~\ref{fig:7}(a)), the magnitudes of $Q_h$, $Q_c$, and the produced work remain relatively moderate. In contrast, when the temperature gradient increases to $T_h=4T_c$ (Fig.~\ref{fig:7}(b)), both the absorbed heat and the produced work become significantly larger. This behavior is expected since a larger temperature difference enhances the energy exchange between the system and the reservoirs, leading to an improvement in the thermodynamic output of the cycle.

From a metrological viewpoint, this behavior is closely related to the parameters previously estimated through the quantum Fisher information analysis. In particular, the parameter $\omega_B$ controls the energy gap of the system and therefore determines both the thermal population distribution and the sensitivity of the system to parameter variations. The optimal estimation regions obtained in the metrological study thus correspond to favorable working regimes for the quantum Stirling cycle, establishing a direct connection between quantum parameter estimation and the thermodynamic performance of the gravcat system.
\begin{figure}[H]
	\centering
	\includegraphics[scale=0.8]{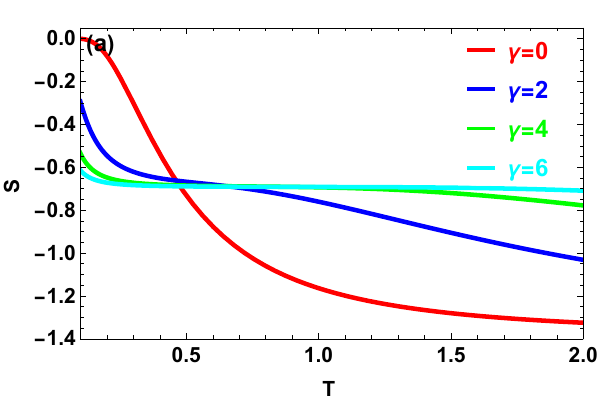}
	\hspace*{0.3cm}
	\includegraphics[scale=0.8]{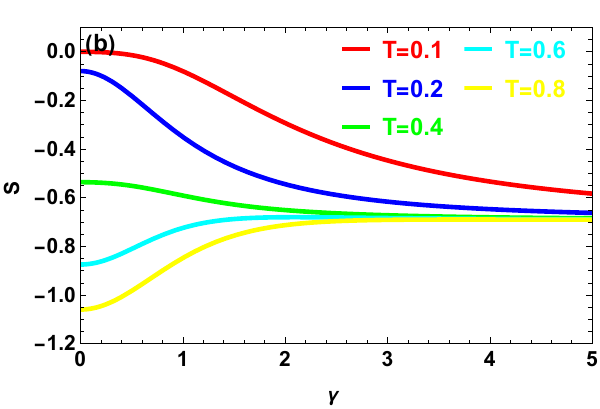}
	\caption{Variation of the entropy as a function of (a) the temperature $T$ for different values of $\gamma$ with $\omega=1$, and (b) the parameter $\gamma$ for different temperatures $T$, with $\omega=1$.}
	\label{fig:8}
\end{figure}
Fig.~\ref{fig:8}(a) presents the variation of the von Neumann entropy $S$ as a function of the temperature $T$ for several values of the coupling parameter $\gamma$ with $\omega=1$. As the temperature increases, the entropy generally decreases, indicating a gradual modification of the population distribution among the energy levels of the system. The influence of the coupling parameter is also evident: larger values of $\gamma$ tend to smooth the temperature dependence of the entropy, reflecting the role of the interaction strength in modifying the energy spectrum and consequently the thermodynamic properties of the system. 

The dependence of the entropy on the coupling parameter $\gamma$ for different temperatures is shown in Fig.~\ref{fig:8}(b). For small values of $\gamma$, the entropy exhibits a stronger sensitivity to temperature variations, while for larger values of the coupling parameter the curves tend to converge toward similar values. This behavior indicates that strong interaction regimes reduce the thermal sensitivity of the system, leading to a more stable entropy profile. Such behavior highlights the interplay between thermal effects and the gravitational coupling in determining the thermodynamic properties of the gravcat system. 
\begin{figure}[H]
	\centering
	\includegraphics[scale=0.8]{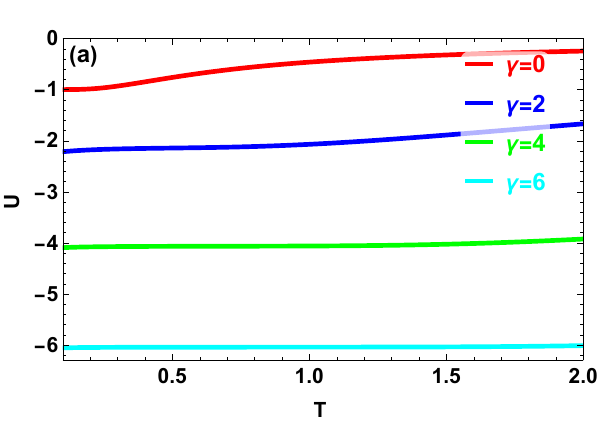}
	\hspace*{0.3cm}
	\includegraphics[scale=0.8]{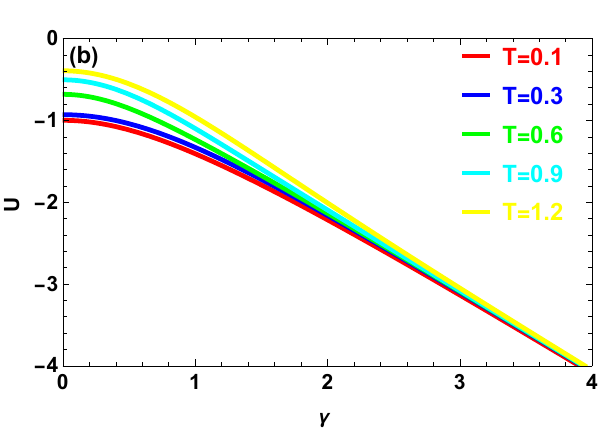}
	\caption{Variation of the internal energy as a function of (a) the temperature $T$ for different values of $\gamma$ with $\omega=1$, and (b) the parameter $\gamma$ for different temperatures $T$, with $\omega=1$.}
	\label{fig:9}
\end{figure}
Fig.~\ref{fig:9}(a) illustrates the variation of the internal energy $U$ as a function of the temperature $T$ for several values of the coupling parameter $\gamma$ with $\omega=1$. The results show that the internal energy increases progressively with temperature for all considered values of $\gamma$. This behavior is expected since higher temperatures lead to a larger population of the excited energy levels, which increases the average energy of the system. In addition, increasing the coupling parameter shifts the internal energy toward more negative values, indicating that stronger interaction modifies the structure of the energy spectrum and therefore the thermodynamic state of the system. 

The dependence of the internal energy on the coupling parameter $\gamma$ for different temperatures is presented in Fig.~\ref{fig:9}(b). The curves indicate that the internal energy decreases as the interaction strength increases, while higher temperatures lead to larger energy values. This behavior highlights the combined effect of thermal excitations and interaction strength in determining the energy content of the gravcat system. These results confirm that both the temperature and the coupling parameter play a significant role in shaping the thermodynamic properties of the system. 
\begin{figure}[H]
	\centering
	\includegraphics[scale=0.8]{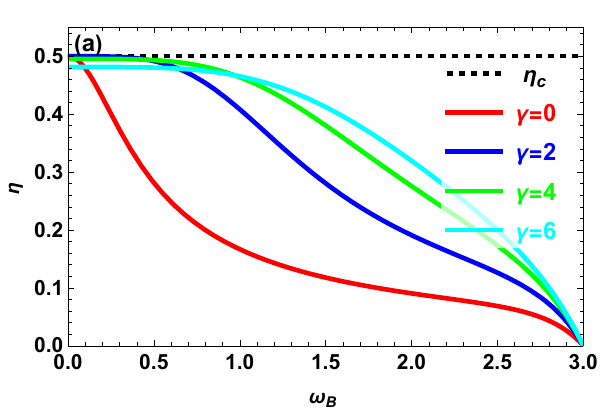}
	\hspace*{0.3cm}
	\includegraphics[scale=0.8]{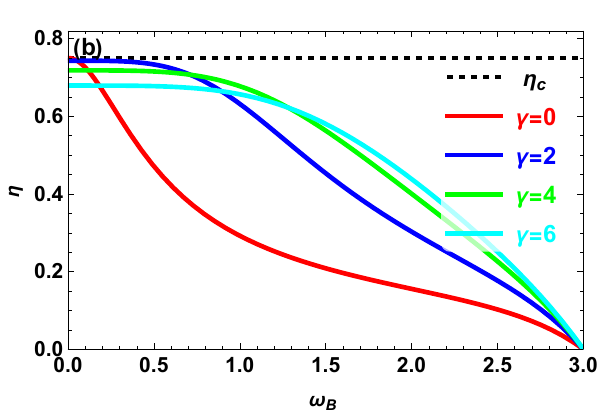}
\caption{Variation of the efficiency as a function of $\omega_B$ for different values of $\gamma$ with $\omega_A=3$: 
	(a) $T_h=2T_c$; (b) $T_h=4T_c$. The Carnot efficiency is given by $\eta_c = 1-{T_c}/{T_h}$.}
	\label{fig:10}
\end{figure}
Fig.~\ref{fig:10} shows the variation of the efficiency $\eta$ of the quantum Stirling engine as a function of the parameter $\omega_B$ for several values of the coupling parameter $\gamma$, with $\omega_A=3$. The dashed horizontal line represents the Carnot efficiency $\eta_c=1-{T_c}/{T_h}$, which sets the theoretical upper bound for the efficiency of the heat engine.

In the case $T_h=2T_c$ (Fig.~\ref{fig:10}(a)), the efficiency initially takes values close to the Carnot limit for small values of $\omega_B$, indicating that the engine operates near its optimal thermodynamic regime. As $\omega_B$ increases, the efficiency gradually decreases and eventually approaches zero, reflecting the reduction of useful work extracted from the cycle. The influence of the coupling parameter $\gamma$ is also evident, since larger values of $\gamma$ tend to maintain higher efficiency over a wider range of $\omega_B$.

When the temperature difference increases to $T_h=4T_c$ (Fig.~\ref{fig:10}(b)), the overall efficiency becomes significantly larger, approaching a higher Carnot bound. In this regime, the curves remain close to the Carnot efficiency for small $\omega_B$, demonstrating that a stronger thermal gradient enhances the thermodynamic performance of the engine. However, as in the previous case, increasing $\omega_B$ reduces the efficiency due to the modification of the system energy spectrum and the corresponding decrease in the extracted work. These results highlight the combined role of the interaction strength and the temperature gradient in determining the efficiency of the gravcat-based quantum Stirling engine.

\section{Conclusion}\label{sec7}

In summary, we have investigated the multiparameter quantum estimation and quantum thermodynamic properties of a gravitational cat state (gravcat) system composed of two interacting massive particles confined in double-well potentials. The system was described through an effective Hamiltonian that incorporates both the energy splitting parameter $\omega$ and the gravitational coupling strength $\gamma$, while the interaction with a thermal environment was taken into account through the Gibbs thermal state. Using the framework of the quantum Fisher information matrix (QFIM), we analyzed the estimation precision of the three fundamental parameters characterizing the model, namely the gravitational coupling $\gamma$, the energy splitting $\omega$, and the temperature $T$. Both simultaneous and individual estimation strategies were examined. Our results showed that the achievable precision strongly depended on the thermal regime as well as on the values of the system parameters. In particular, the existence of optimal estimation regions was identified for different parameter pairs, where the minimal variances reach their lowest values. The comparison between the simultaneous and individual estimation protocols revealed that the relative performance of the two strategies depends on the interplay between temperature, interaction strength, and energy gap. In several parameter regions, the simultaneous estimation scheme provides higher precision, whereas in other regimes the individual estimation may become more advantageous. These results highlight the important role of multiparameter quantum estimation in extracting physical information from gravcat systems. Furthermore, we investigated the thermodynamic behavior of the system within the framework of a quantum Stirling cycle. The internal energy, entropy, heat exchanges, and work production were analyzed, allowing us to evaluate the performance of the gravcat system as a quantum heat engine. The results show that the thermodynamic quantities and the engine efficiency are strongly influenced by the system parameters and the temperature gradient between the thermal reservoirs. Overall, this study demonstrates that gravitationally interacting quantum systems can provide a promising platform for exploring the interplay between quantum metrology and quantum thermodynamics. The estimation of optimal parameter regimes not only enhances the precision of quantum measurements but also provides useful information for optimizing the thermodynamic performance of quantum heat engines based on gravcat systems.
\appendix
\section{Explicit expressions of the QFIM elements}\label{Appendix A}
\subsection{QFIM for the estimation of $\gamma$ and $T$}

The Quantum Fisher Information Matrix corresponding to the parameters
$\xi=(\gamma,T)$ is written as

\begin{equation}
	\mathcal{F}=
	\begin{pmatrix}
		\mathcal{F}_{\gamma\gamma} & \mathcal{F}_{\gamma T} \\
		\mathcal{F}_{T\gamma} & \mathcal{F}_{TT}
	\end{pmatrix}.
\end{equation}
For simplicity, we introduce the quantity
$
\Delta = \sqrt{\gamma^2+\omega^2}
$.
The explicit expressions for the elements of $\mathcal{F}$ are given by

\begin{align}
	\mathcal{F}_{\gamma\gamma} &=
	-\frac{\text{sech}\left(\frac{\Delta}{T}\right)}
	{8T^2\Delta^4
		\left(\cosh\left(\frac{\Delta}{T}\right)
		+\cosh\left(\frac{\gamma}{T}\right)\right)^2
		\left(-5\gamma^2+3\gamma^2\cosh\left(\frac{2\Delta}{T}\right)-2\omega^2\right)}
	\nonumber \\
	&\times \Bigg[
	2\cosh\left(\frac{\gamma}{T}\right)
	\Big(
	26\gamma^6
	+\omega^4(19\gamma^2-16T^2)
	+\gamma^2\omega^2(41\gamma^2-13T^2)
	-3\gamma^2\left(2\gamma^4+\omega^2(3\gamma^2+T^2)+\omega^4\right)
	\cosh\left(\frac{4\Delta}{T}\right)
	\nonumber \\
	&\qquad
	+4\Delta^2
	\left(5\gamma^4+\omega^2(3\gamma^2+4T^2)+\omega^4\right)
	\cosh\left(\frac{2\Delta}{T}\right)
	+4\omega^6
	\Big)
	-2\cosh\left(\frac{\Delta}{T}\right)
	\Big(
	-52\gamma^6-88\gamma^4\omega^2-44\gamma^2\omega^4
	+13\gamma^2T^2\omega^2
	\Big)
	\nonumber \\
	&\qquad
	+4\Delta^2
	\left(3\gamma^4+\omega^2(3\gamma^2-4T^2)\right)
	\cosh\left(\frac{2\Delta}{T}\right)
	+16T^2\omega^4-8\omega^6
	\Bigg],
\end{align}

\begin{align}\nonumber
	\mathcal{F}_{T\gamma}=\mathcal{F}_{\gamma T} &=
	\frac{1}{
		T^3\Delta
		\left(\cosh\left(\frac{\Delta}{T}\right)
		+\cosh\left(\frac{\gamma}{T}\right)\right)^2
		\left(-5\gamma^2+3\gamma^2\cosh\left(\frac{2\Delta}{T}\right)-2\omega^2\right)
	}
	\\\nonumber
	&\times
	\Bigg[
	\gamma\Delta(13\gamma^2+4\omega^2)
	+3\gamma T\omega^2\sinh\left(\frac{2\Delta}{T}\right)
	+\gamma\cosh\left(\frac{\gamma}{T}\right)
	\Big(
	6T\omega^2\sinh\left(\frac{\Delta}{T}\right)
	-3\gamma^2\Delta\cosh\left(\frac{3\Delta}{T}\right)
	\\\nonumber
	&\qquad
	+\Delta(13\gamma^2+4\omega^2)
	\cosh\left(\frac{\Delta}{T}\right)
	\Big)
	+(2\gamma^2+\omega^2)
	\sinh\left(\frac{\gamma}{T}\right)
	\sinh\left(\frac{\Delta}{T}\right)
	\left(-5\gamma^2+3\gamma^2
	\cosh\left(\frac{2\Delta}{T}\right)-2\omega^2\right)
	\\
	&\qquad
	-3\gamma^3\Delta
	\cosh\left(\frac{2\Delta}{T}\right)
	\Bigg],
\end{align}

	\begin{align}\nonumber
		\mathcal{F}_{TT} &=
		\frac{1}{
			T^4\left(\cosh\left(\frac{\Delta}{T}\right)
			+\cosh\left(\frac{\gamma}{T}\right)\right)^2
			\left(-5\gamma^2+3\gamma^2
			\cosh\left(\frac{2\Delta}{T}\right)
			-2\omega^2\right)
		}
		\\\nonumber
		&\times
		\Bigg[
		-13\gamma^4-12\gamma^2\omega^2
		+\cosh\left(\frac{\gamma}{T}\right)
		\cosh\left(\frac{\Delta}{T}\right)
		\Big(
		-16\gamma^4-15\gamma^2\omega^2
		+3\gamma^2(2\gamma^2+\omega^2)
		\cosh\left(\frac{2\Delta}{T}\right)
		-2\omega^4
		\Big)
		\\\nonumber
		&\qquad
		+\gamma\sinh\left(\frac{\Delta}{T}\right)
		\Big(
		\Delta\sinh\left(\frac{\gamma}{T}\right)
		\left(
		13\gamma^2
		-3\gamma^2\sinh^2\left(\frac{\Delta}{T}\right)
		+4\omega^2
		\right)
		+3\gamma^3\sinh\left(\frac{\Delta}{T}\right)
		\Big)
		\\
		&\qquad
		+3\gamma^3
		\left(
		\gamma
		-3\Delta
		\sinh\left(\frac{\gamma}{T}\right)
		\sinh\left(\frac{\Delta}{T}\right)
		\right)
		\cosh^2\left(\frac{\Delta}{T}\right)
		-2\omega^4
		\Bigg].
	\end{align}
	
\subsection{QFIM for the estimation of $\omega$ and $T$}
	The Quantum Fisher Information Matrix (QFIM) associated with the parameter vector $\xi=(\omega,T)$ is defined as
	\begin{equation}
		\mathcal{F}=
		\begin{pmatrix}
			\mathcal{F}_{\omega\omega} & \mathcal{F}_{\omega T} \\
			\mathcal{F}_{T\omega} & \mathcal{F}_{TT}
		\end{pmatrix},
	\end{equation}
	
where

\begin{align}
	\mathcal{F}_{\omega\omega}=
	\frac{\mathcal{N}}
	{8T^2\Delta^{13}
		\left(\cosh\!\left(\frac{\gamma}{T}\right)+\cosh\!\left(\frac{\Delta}{T}\right)\right)^4
		\left(-3\gamma^2\cosh\!\left(\frac{2\Delta}{T}\right)+5\gamma^2+2\omega^2\right)},
\end{align}
with
\begin{align}
	\mathcal{N}=&
	8\omega^2\cosh\!\left(\frac{\gamma}{T}\right)
	\left(\cosh\!\left(\frac{\gamma}{T}\right)+\cosh\!\left(\frac{\Delta}{T}\right)\right)
	\left(-3\gamma^2\cosh\!\left(\frac{2\Delta}{T}\right)+5\gamma^2+2\omega^2\right)
	\sinh^2\!\left(\frac{\Delta}{T}\right)\Delta^{11}
	\nonumber\\
	&+32\gamma^2\omega^2
	\left(\cosh\!\left(\frac{\gamma}{T}\right)+\cosh\!\left(\frac{\Delta}{T}\right)\right)^2
	\text{sech}\!\left(\frac{\Delta}{T}\right)
	\left(\Delta\cosh\!\left(\frac{\Delta}{T}\right)-T\sinh\!\left(\frac{\Delta}{T}\right)\right)
	\nonumber\\
	&\times
	\left(
	2\cosh\!\left(\frac{\gamma}{T}\right)
	\left(\Delta\cosh\!\left(\frac{\Delta}{T}\right)-T\sinh\!\left(\frac{\Delta}{T}\right)\right)
	-T\sinh\!\left(\frac{2\Delta}{T}\right)+2\Delta
	\right)\Delta^{9}
	\nonumber\\
	&+\frac{1}{2}
	\left(\cosh\!\left(\frac{\gamma}{T}\right)\text{sech}\!\left(\frac{\Delta}{T}\right)+1\right)
	\Big[
	6T\omega\gamma^2
	\cosh\!\left(\frac{\Delta}{T}\right)
	\cosh\!\left(\frac{3\Delta}{T}\right)
	+2\omega\left(4\omega(\gamma^2+\omega^2)-3T\gamma^2\right)
	\cosh^2\!\left(\frac{\Delta}{T}\right)
	\nonumber\\
	&\quad
	+4\omega\cosh\!\left(\frac{\gamma}{T}\right)
	\cosh\!\left(\frac{\Delta}{T}\right)
	\left(-3T\gamma^2+3T\gamma^2\cosh\!\left(\frac{2\Delta}{T}\right)+2\omega(\gamma^2+\omega^2)\right)
	\nonumber\\
	&\quad
	-2\Delta
	\Big(
	2\cosh\!\left(\frac{\gamma}{T}\right)
	\left(-5T+3\omega+3(T+\omega)\cosh\!\left(\frac{2\Delta}{T}\right)\right)\gamma^2
	\nonumber\\
	&\qquad
	+2\cosh\!\left(\frac{\Delta}{T}\right)
	\left(2\omega^3+8\gamma^2\omega-5T\gamma^2
	+3T\gamma^2\cosh\!\left(\frac{2\Delta}{T}\right)\right)
	\Big)
	\sinh\!\left(\frac{\Delta}{T}\right)
	\Big],
\end{align}

\begin{align}\nonumber
	\mathcal{F}_{\omega T} =	\mathcal{F}_{T\omega}&= \frac{\omega}{2 T^3 \Delta \left(\cosh \left(\frac{\Delta}{T}\right) + \cosh \left(\frac{\gamma}{T}\right)\right)^2 \left(5 \gamma^2 - 3 \gamma^2 \cosh \left(\frac{2 \Delta}{T}\right) + 2 \omega^2\right)} \\
	\nonumber
	&\Biggl[ -4 \Delta (4 \gamma^2 + \omega^2) + 6 \gamma^2 T \sinh \left(\frac{2 \Delta}{T}\right) 
	+ 2 \gamma \sinh \left(\frac{\gamma}{T}\right) \sinh \left(\frac{\Delta}{T}\right) \left(5 \gamma^2 - 3 \gamma^2 \cosh \left(\frac{2 \Delta}{T}\right) + 2 \omega^2\right) \\
	&+ \cosh \left(\frac{\gamma}{T}\right) \biggl( 3 \gamma^2 \left(4 T \sinh \left(\frac{\Delta}{T}\right) + \Delta \cosh \left(\frac{3 \Delta}{T}\right)\right) - \Delta (19 \gamma^2 + 4 \omega^2) \cosh \left(\frac{\Delta}{T}\right) \biggr) \Biggr].
\end{align}

\subsection{QFIM for the estimation of $\omega$ and $\gamma$}

The Quantum Fisher Information Matrix (QFIM) for the parameter vector $\xi=(\omega,\gamma)$ takes the form
\begin{equation}
	\mathcal{F}=
	\begin{pmatrix}
		\mathcal{F}_{\omega\omega} & \mathcal{F}_{\omega \gamma} \\
		\mathcal{F}_{\gamma\omega} & \mathcal{F}_{\gamma\gamma}
	\end{pmatrix},
\end{equation}
with
\begin{align}\nonumber
	\mathcal{F}_{\omega \gamma}  &= \frac{1}{4 T^2 \Delta^{11} \left(\cosh \left(\frac{\gamma}{T}\right)+\cosh \left(\frac{\Delta}{T}\right)\right)^4 \left(5 \gamma^2+2 \omega^2-3 \gamma^2 \cosh \left(\frac{2 \Delta}{T}\right)\right)} \Biggl[ \\\nonumber
	&-4 \omega \Delta^9 \left(\cosh \frac{\gamma}{T}+\cosh \frac{\Delta}{T}\right) \left(5 \gamma^2+2 \omega^2-3 \gamma^2 \cosh \frac{2 \Delta}{T}\right) \sinh \frac{\Delta}{T} \left(\Delta \cosh \frac{\Delta}{T} \sinh \frac{\gamma}{T}-\gamma \cosh \frac{\gamma}{T} \sinh \frac{\Delta}{T}\right) \\\nonumber
	&-16 \gamma \omega \Delta^7 \left(\cosh \frac{\gamma}{T}+\cosh \frac{\Delta}{T}\right)^2 \operatorname{sech} \frac{\Delta}{T} \left(T \sinh \frac{\Delta}{T}-\Delta \cosh \frac{\Delta}{T}\right) \biggl(2 \Delta \gamma^2 \cosh \frac{\gamma}{T} \cosh \frac{\Delta}{T} + 2 \Delta \gamma^2 \\\nonumber
	&+ 2 \left(T \omega^2 \cosh \frac{\gamma}{T}-\gamma \Delta^2 \sinh \frac{\gamma}{T}\right) \sinh \frac{\Delta}{T} + T \omega^2 \sinh \frac{2 \Delta}{T}\biggr) \\\nonumber
	&+ \frac{1}{2} \left(\frac{\cosh (\gamma/T)}{\cosh (\Delta/T)}+1\right) \biggl(6 T \omega \cosh \frac{\Delta}{T} \cosh \frac{3 \Delta}{T} \gamma^2 + 2 \omega \left(4 \omega \Delta^2-3 T \gamma^2\right) \cosh^2 \frac{\Delta}{T} \\\nonumber
	&+ 4 \omega \cosh \frac{\gamma}{T} \cosh \frac{\Delta}{T} \left(-3 T \gamma^2+3 T \gamma^2 \cosh \frac{2 \Delta}{T}+2 \omega \Delta^2\right) \\\nonumber
	&- 2 \Delta \left(2 \cosh \frac{\gamma}{T} \left(-5 T+3 \omega+3(T+\omega) \cosh \frac{2 \Delta}{T}\right) \gamma^2 + 2 \cosh \frac{\Delta}{T} \left(2 \omega^3+8 \gamma^2 \omega-5 T \gamma^2+3 T \gamma^2 \cosh \frac{2 \Delta}{T}\right)\right) \sinh \frac{\Delta}{T}\biggr) \\\nonumber
	&\times \left(-\Delta^3 \cosh \frac{\Delta}{T} \sinh \frac{\gamma}{T} - \omega \Delta^2 \sinh \frac{\gamma}{T} \sinh \frac{\Delta}{T} + \gamma \omega \Delta + \gamma \cosh \frac{\gamma}{T} \left(\omega \Delta \cosh \frac{\Delta}{T} + (\Delta^2-T \omega) \sinh \frac{\Delta}{T}\right) - \frac{1}{2} T \gamma \omega \sinh \frac{2 \Delta}{T}\right) \Delta^6 \\
	\Biggr],
\end{align}
\begin{align}\nonumber
\mathcal{F}_{\gamma\omega}= & \frac{1}{4 T^2 \Delta^{11} \left(\cosh \left(\frac{\gamma}{T}\right)+\cosh \left(\frac{\Delta}{T}\right)\right)^4 \left(3 \gamma^2 \cosh \left(\frac{2 \Delta}{T}\right)-5 \gamma^2-2 \omega^2\right)} \Biggl[ \\\nonumber
	& -4 \omega \Delta^{10} \left(\cosh \left(\frac{\gamma}{T}\right)+\cosh \left(\frac{\Delta}{T}\right)\right)^2 \left(3 \gamma^2 \cosh \left(\frac{2 \Delta}{T}\right)-5 \gamma^2-2 \omega^2\right) \sinh \left(\frac{\gamma}{T}\right) \sinh \left(\frac{\Delta}{T}\right) \\\nonumber
	& + 4 \omega \Delta^9 \cosh \left(\frac{\gamma}{T}\right) \left(\cosh \left(\frac{\gamma}{T}\right)+\cosh \left(\frac{\Delta}{T}\right)\right) \left(-3 \gamma^2 \cosh \left(\frac{2 \Delta}{T}\right)+5 \gamma^2+2 \omega^2\right) \sinh \left(\frac{\Delta}{T}\right) \left(\Delta \sinh \frac{\gamma}{T}+\gamma \sinh \frac{\Delta}{T}\right) \\\nonumber
	& + 16 \gamma \omega \Delta^7 \left(\cosh \frac{\gamma}{T}+\cosh \frac{\Delta}{T}\right)^2 \operatorname{sech} \frac{\Delta}{T} \left(\Delta \gamma^2 \cosh \frac{\Delta}{T}+T \omega^2 \sinh \frac{\Delta}{T}\right) \times \left(2 \cosh \frac{\gamma}{T} \left(\Delta \cosh \frac{\Delta}{T}-T \sinh \frac{\Delta}{T}\right)-T \sinh \frac{2 \Delta}{T}+2 \Delta\right) \\\nonumber
	& + \frac{\Delta^4}{2} \left(T \gamma^2 \sinh \frac{2 \Delta}{T}+2 \cosh \frac{\gamma}{T} \left(\Delta \omega^2 \cosh \frac{\Delta}{T}+\left(\omega^3+\gamma^2 \omega+T \gamma^2\right) \sinh \frac{\Delta}{T}\right)+2 \omega^2 \Delta\right) \\\nonumber
	& \quad \times \biggl(2 \Delta^4 \cosh \frac{\Delta}{T} \left(3 \gamma^2 \cosh \frac{2 \Delta}{T}-5 \gamma^2-2 \omega^2\right) \sinh \frac{\gamma}{T} - \gamma \Delta^3 \left(8 \gamma^2+\omega(5 T+2 \omega)-3 T \omega \cosh \frac{2 \Delta}{T}\right) \sinh \frac{2 \Delta}{T} \\\nonumber
	& \quad + 2 \gamma \omega \Delta^2 \cosh^2 \frac{\Delta}{T} \left(2 \gamma^2+\omega(3 T+2 \omega)-3 T \omega \cosh \frac{2 \Delta}{T}\right) \\\nonumber
	& \quad - \Delta^2 \cosh \frac{\gamma}{T} \Bigl( -4 \gamma \omega \cosh \frac{\Delta}{T} \left(2 \Delta^2+3 T \omega-3 T \omega \cosh \frac{2 \Delta}{T}\right) - 2 \Delta^2 \left(3 \gamma^2 \cosh \frac{2 \Delta}{T}-5 \gamma^2-2 \omega^2\right) \sinh \frac{\gamma}{T} \\\nonumber
	& \quad \quad + 2 \gamma \Delta \left(11 \gamma^2+2 \omega(5 T+\omega)+3\left(\gamma^2-2 T \omega\right) \cosh \frac{2 \Delta}{T}\right) \sinh \frac{\Delta}{T} \Bigr) \\\nonumber
	& \quad + \gamma \Delta^2 \cosh^2 \frac{\gamma}{T} \Bigl( -6 T \omega^2 \cosh \frac{2 \Delta}{T}+2 \omega\left(2 \gamma^2+\omega(3 T+2 \omega)\right)  \quad - 2 \Delta \left(3 \gamma^2+5 T \omega+3\left(\gamma^2-T \omega\right) \cosh \frac{2 \Delta}{T}\right) \tanh \frac{\Delta}{T} \Bigr) \biggr) \\\nonumber
	& + 2 \Delta \left(\omega^4+\gamma^2 \omega^2+\cosh \frac{\gamma}{T} \left(\omega^2 \Delta^2 \cosh \frac{\Delta}{T}-\Delta (\omega^3+\gamma^2 \omega-T \gamma^2) \sinh \frac{\Delta}{T}\right)+\frac{1}{2} T \gamma^2 \Delta \sinh \frac{2 \Delta}{T}\right) \\\nonumber
	& \quad \times \biggl( -4 \gamma^2 \Delta^2 \left(\cosh \frac{\gamma}{T}+\cosh \frac{\Delta}{T}\right) \sinh \frac{\Delta}{T} \left[ -\Delta^3 \cosh \frac{\Delta}{T} \sinh \frac{\gamma}{T} - \omega \Delta^2 \sinh \frac{\gamma}{T} \sinh \frac{\Delta}{T} + \gamma \omega \Delta \right. \\\nonumber
	& \quad \quad \left. + \gamma \cosh \frac{\gamma}{T} \left(\omega \Delta \cosh \frac{\Delta}{T} + (\Delta^2 - T \omega) \sinh \frac{\Delta}{T}\right) - \frac{1}{2} T \gamma \omega \sinh \frac{2 \Delta}{T} \right] \tanh \frac{\Delta}{T} \\\nonumber
	& \quad + 4 \gamma \Delta \left(\cosh \frac{\gamma}{T} + \cosh \frac{\Delta}{T}\right) \left(\Delta^2 \cosh \frac{\Delta}{T} + \omega \Delta \sinh \frac{\Delta}{T}\right) \tanh \frac{\Delta}{T} \\\nonumber
	& \quad \quad \times \left(2 \gamma^2 \Delta \cosh \frac{\gamma}{T} \cosh \frac{\Delta}{T} + 2 \gamma^2 \Delta + 2 (T \omega^2 \cosh \frac{\gamma}{T} - \gamma \Delta^2 \sinh \frac{\gamma}{T}) \sinh \frac{\Delta}{T} + T \omega^2 \sinh \frac{2 \Delta}{T}\right) \\
	& \quad + \left(\frac{\cosh (\gamma/T)}{\cosh (\Delta/T)} + 1\right) \left[ \Delta^4 \cosh \frac{\Delta}{T} \sinh \frac{\gamma}{T} + \gamma \omega \Delta^2 - \omega \Delta \left(T \gamma \cosh \frac{\Delta}{T} + \Delta^2 \sinh \frac{\gamma}{T}\right) \sinh \frac{\Delta}{T} \right. \\\nonumber
	& \quad \quad \left. - \gamma \cosh \frac{\gamma}{T} \left(\Delta (\gamma^2 + \omega (T + \omega)) \sinh \frac{\Delta}{T} - \omega \Delta^2 \cosh \frac{\Delta}{T}\right) \right] \times \left( \Delta (3 \gamma^2 + \omega^2) + (\omega^2 - \gamma^2) \Delta \cosh \frac{2 \Delta}{T} + \omega \Delta^2 \sinh \frac{2 \Delta}{T} \right) \biggr) \Biggr].
\end{align}

\end{document}